\documentclass{article}

\usepackage[preprint]{neurips_2025}
\usepackage[utf8]{inputenc} 
\usepackage[T1]{fontenc}    
\usepackage[colorlinks=true, linkcolor=blue, citecolor=blue, urlcolor=blue]{hyperref}
\usepackage{url}            
\usepackage{booktabs}       
\usepackage{amsfonts}       
\usepackage{nicefrac}       
\usepackage{microtype}      
\usepackage{xcolor}         
\usepackage{subcaption}
\usepackage{booktabs}
\usepackage{graphicx}
\usepackage{amsmath}
\usepackage{placeins}
\RequirePackage{silence}
\usepackage[capitalize,noabbrev]{cleveref}

\crefname{section}{Section}{Sections}
\crefname{equation}{Eq.}{Eqs.}
\crefname{figure}{Figure}{Figures}
\crefname{table}{Table}{Tables}

\title{Structured SIR: Efficient and Expressive Importance-Weighted Inference for High-Dimensional Image Registration}

\author{Ivor J. A. Simpson\\
Department of Informatics \\
University of Sussex, UK\\
  \texttt{ i.simpson@sussex.ac.uk} \\
\And
Neill D. F. Campbell\\
Department of Computer Science \\ 
University College London, UK \\ \\
Department of Computer Science \\
University of Bath, UK\\
  \texttt{neill.campbell@ucl.ac.uk}
}

\begin{document}
\maketitle
\newcommand{\qi}{\tilde{q}}
\newcommand{\qv}[1]{q_\psi(#1)}

\newcommand{\Ifix}{\mathbf{I}_f}
\newcommand{\Imov}{\mathbf{I}_m}

\newcommand{\warpp}{\mathbf{Z}}

\newcommand{\qvwarp}{\qv{\warpp}}

\newcommand{\qmean}{\boldsymbol{\mu}}
\newcommand{\qcov}{\boldsymbol{\Sigma}}
\newcommand{\vecf}{\text{vec}}

\newcommand{\Lmat}{\mathbf{L}}
\newcommand{\Rmat}{\mathbf{R}}

\newcommand{\vecwarpp}{\vecf(\warpp)}
\newcommand{\xb}{\mathbf{x}}
\newcommand{\A}{\mathbf{A}}
\newcommand{\U}{\mathbf{U}}
\newcommand{\V}{\mathbf{V}}
\newcommand{\C}{\mathbf{C}}
\newcommand{\M}{\mathbf{M}}
\newcommand{\kb}{\mathbf{k}}
\newcommand{\Sb}{\mathbf{S}}

\begin{abstract}
  Image registration is an ill-posed dense vision task, where multiple  solutions achieve similar loss values, motivating probabilistic inference. Variational inference has previously been employed to capture these distributions, however restrictive assumptions about the posterior form can lead to poor characterisation, overconfidence and low-quality samples. More flexible posteriors are typically bottlenecked by the complexity of high-dimensional covariance matrices required for dense 3D image registration.

In this work, we present a memory and computationally efficient inference method, Structured SIR, that enables expressive, multi-modal, characterisation of uncertainty with high quality samples. We propose the use of a Sampled Importance Resampling (SIR) algorithm with a novel memory-efficient high-dimensional covariance parameterisation as the sum of a low-rank covariance and a sparse, spatially structured Cholesky precision factor. This structure enables capturing complex spatial correlations while remaining computationally tractable.

We evaluate the efficacy of this approach in 3D dense image registration of brain MRI data, which is a very high-dimensional problem. We demonstrate that our proposed method produces 
uncertainty estimates that are significantly better calibrated than those produced by variational methods, achieving equivalent or better accuracy. Crucially, we show that the model yields highly structured multi-modal posterior distributions, enable effective and efficient uncertainty quantification.

\end{abstract}
\section{Introduction}
Dense image registration, establishing pixel-, or voxel-wise geometric correspondences between images, is a fundamental task in computer vision and medical image analysis. 
There is intrinsic ambiguity in whether or not images are matched so this
problem is fundamentally ill-posed; there will be multiple plausible registration results. This has lead to the use of probabilistic inference approaches both without  \citep{le2016quantifying,risholm2011multimodal,simpson2015probabilistic} and with the assistance of neural networks \citep{dalca2019unsupervised,grzech2021uncertainty,siegert2024pulpo}.
The use of deep learning has provided some benefits for medical image registration, particularly with respect to computational efficiency via amortised inference~\citep{dalca2019unsupervised}. 
Nonetheless, the expressivity of the posterior distributions provided by existing approaches is limited by the memory scaling of high-dimensional covariance matrices. Moreover, many approaches follow a variational paradigm with the strong assumption that the posterior is an (often factorised) Gaussian, making it impossible to capture multi-modal outputs.

In this work, we propose a framework that relaxes the limitations of variational inference and simplistic posterior distributions. Our contributions include:
\begin{enumerate}
    \item Introducing an amortised Importance Sampling procedure, Sampled Importance Resampling (SIR), to produce a \textit{proposal distribution}, from which samples can be efficiently drawn and weighted, to establish a more flexible posterior than variational inference that matches the true posterior.
    \item Providing an expressive Gaussian formulation for our proposal distribution, with a novel memory-efficient covariance parameterisation, consisting of a sum of a low-rank covariance and a spatially structured locally sparse Cholesky factor of the precision, with cross-directional correlations.
    \item Demonstrating the improvements of our inferred importance weighted registration distributions, when compared to variational equivalents, in terms of: (i)~segmentation accuracy; (ii)~quality and coherence of samples; and, importantly, (iii)~probabilistic calibration (i.e.~the predicted uncertainty correlates with registration error) on a brain MRI dataset.
    
\end{enumerate}

\section{Background}
\subsection{Registration Notation}
We denote the fixed and moving images as $\Ifix$ and $\Imov$, defined over a common discrete spatial domain $\Omega$ containing $N_v$ voxels; that is, we are learning the deformation that takes $\Imov$ to $\Ifix$. 
The dense displacement field is stacked in matrix $\warpp \in \mathbb{R}^{3\times N_{v}}$, where any column $\mathbf{z} \in \mathbb{R}^3$ is a 3D displacement. The warped image is given by $\Imov(\mathbf{x} + \mathbf{z}(\mathbf{x}))$, denoted as $\Imov \circ \warpp$ for brevity.

\subsection{Probabilistic Registration}
While traditional registration provides a single MAP (Maximum-A-Posteriori) estimate, e.g.~\citep{ashburner2007fast}, fully probabilistic frameworks aim to characterise the  posterior distribution $p(\warpp | \Ifix, \Imov)$. This is particularly critical for medical applications, where confidence intervals can guide clinical decision making. 
However, Uncertainty Quantification (UQ) has been challenging for registration, in general, and for many approaches the characterised uncertainty has been found to correlate poorly with the needs of downstream applications~\citep{luo2019applicability}; we seek to rectify this situation with well calibrated uncertainty and coherent posterior samples.

\subsubsection{Sampling and Approximate Inference:} Previous works attempt to capture the posterior distribution using either: sampling based methods \citep{risholm2011multimodal,zhang2013bayesian}, which tend to be slow for dense 3D registration; or variational~\citep{le2016quantifying,simpson2015probabilistic}, or Laplace approximation~\citep{wang2018efficient}, approaches that are both computationally intensive and only provide a unimodal approximation of the posterior.

More recently, Grzech et al.~\citep{grzech2021uncertainty} bridged the gap between deep learning and classical inference by employing Stochastic Gradient MCMC. While this allows for more robust posterior coverage than standard variational methods, the requirement to draw multiple samples at inference time maintains a significant computational overhead compared to single-pass amortized models.

\subsubsection{Amortised Variational Inference:} Amortised methods were introduced to resolve the computational expense of estimating an approximate posterior distribution~\citep{dalca2019unsupervised}. In this framework, a neural network is trained to predict directly the parameters of an approximate posterior $q_{\theta}(\warpp | \Ifix, \Imov)$; typically a multivariate Gaussian. This is achieved by maximizing the Evidence Lower Bound (ELBO)

\begin{align}
    \mathcal{L}_\mathrm{elbo} &= \mathbb{E}_{q_{\theta}(\warpp | \Ifix, \Imov)}\big[ \log p(\Ifix | \Imov \circ \warpp) \big] + \mathrm{KL}\big[ q_{\theta}(\warpp | \Ifix, \Imov) ||p(\warpp) \big]  \,.
    \label{eq:elbo}
\end{align}

\subsubsection{Uncertainty Calibration:} Although amortised variational methods are efficient, their resulting uncertainty estimates are often poorly calibrated, underestimating the posterior variance~\citep{cremer2018inference} and fail to capture multi-modal distributions by construction. To address the multi-scale nature of registration uncertainty, PULPo~\citep{siegert2024pulpo} introduced a hierarchical Laplacian pyramid approach to capture uncertainty across multiple spatial scales. While PULPo improves UQ quality, its iterative/hierarchical nature adds complexity to the sampling procedure.

\subsection{Importance Sampling}
Amortised importance sampling based inference was first introduced in the Importance Weighted Auto Encoder \citep{burda2015importance}; a neural network was used to predict a ``sample proposal distribution'', rather than an approximate posterior, where samples are weighted at test time based on their likelihood, prior and proposal probabilities.  

Importance distributions can be estimated in the same manner as variational methods, where we instead optimise the importance weighting
\begin{equation}
    \mathbb{E}_{z^{1:K}}\left[ \log \left(\frac{1}{K}\sum^{K}_{i=1} \frac{p(\Ifix, \Imov \circ \warpp^i)}{q_{\theta}(\warpp^i | \Ifix, \Imov)}\right)\right] = \mathbb{E}_{z^{1:K}}\left[ \log \left(\frac{1}{K}\sum^{K}_{i=1} \alpha^i \right)\right] \,,
    \label{eq:iwae}
\end{equation}
where
\begin{equation}
    \alpha^i = \frac{p(\Ifix | \Imov \circ \warpp^i) p(\warpp^i)}{q_{\theta}(\warpp^i | \Ifix, \Imov)}
\end{equation}
are the unnormalised weights; superscripts, e.g.~$\cdot^i$, denote samples.

The key distinction from variational inference is that we take the sum of the probabilities, rather than log probabilities, allowing better samples $z^i$ to dominate in the loss (and therefore gradients). The importance log probabilities are thus normalised by the other drawn samples, i.e.~assigned importance weights. We note that using a single sample to calculate the expectation is equivalent to variational inference. 
Previous work demonstrated the multi-modal nature of the inferred distribution\citep{cremer2017reinterpreting} but, to the best of our knowledge, this has not been directly applied to dense imaging problems.

\subsection{Flexible Covariances}
Structured Uncertainty Prediction Networks~\citep{dorta2018structured}, demonstrated an efficient amortised inference approximation to a full-rank covariance via a sparse parameterisation of the \emph{precision} (the inverse of the covariance) equivalent to a Gaussian Markov Random Field~\citep{rue2005gaussian}. 
Low-rank covariance approximations have also been shown to be effectively learned using amortised inference \citep{monteiro2020stochastic}. In this work, we integrate both these formulations together into a more general and flexible parameterisation.

\section{Methods}

\begin{figure}[t]
    \centering
    \begin{subfigure}[t]{0.32\textwidth}
        \centering
        \includegraphics[width=\linewidth]{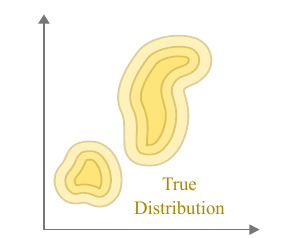}
        \caption{}
        \label{fig:overview_true_dist}
    \end{subfigure}
    \hfill
    \begin{subfigure}[t]{0.32\textwidth}
        \centering
        \includegraphics[width=\linewidth]{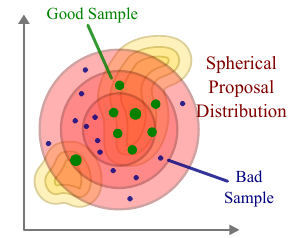}
        \caption{}
        \label{fig:overview_spherical_sampling}
    \end{subfigure}
    \hfill
    \begin{subfigure}[t]{0.32\textwidth}
        \centering
        \includegraphics[width=\linewidth]{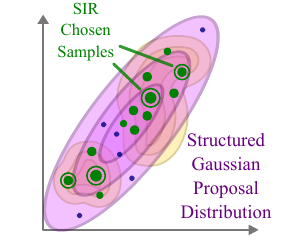}
        \caption{}
        \label{fig:overview_supn_sampling}
    \end{subfigure}
    \caption{Illustration of the posterior modelling problem, where the underlying probability distribution is illustrated in (\subref{fig:overview_true_dist}). (\subref{fig:overview_spherical_sampling})~shows an attempt to model this variationally using a spherical Gaussian, which assigns probability mass to a lot of area that is unlikely, but simultaneously does not cover the whole distribution; if used as a proposal distribution then many poor samples are drawn and this is exacerbated as dimensionality increases. (\subref{fig:overview_supn_sampling})~illustrates how our SIR sampling framework allows selection of high probability samples at both learning/inference time, which avoids penalising a model for being more exploratory. Moreover, using a structured Gaussian allows assigning probability mass to higher density areas improving the efficiency of sampling.}
    \label{fig:overview}
\end{figure}

\subsection{Sampled Importance Resampling (SIR)}
To approximate the complex posterior $p(\warpp | \Ifix, \Imov)$, we employ the SIR framework \citep{rubin1987comment, smith1992bayesian}, a form of importance weighted inference that approximates the objective in \cref{eq:iwae}.  While traditional approaches use all the samples from a proposal distribution to compute a weighted average, we draw inspiration from the resampling-based objectives in \citep{maddison2017filtering}, which focuses the gradient computation on a subset of high-quality samples.

Specifically, we use SIR to identify high-probability registration hypotheses from a large set of candidates drawn from the initial proposal distribution; these selected samples are then used to provide an informative supervisory signal for optimising the parameters of our proposal distribution network. This reduces the overhead of backpropagating through low-weight candidates, and permits a more expansive search of the proposal space. We draw $N_{s}$ samples from the learned proposal distribution $q(\warpp) = \mathcal{N}(\qmean, \qcov)$ using the reparameterisation trick
\begin{align}
    \warpp^i = \qmean + \Rmat \boldsymbol{\epsilon}^i_R + \Lmat^{-\top}\boldsymbol{\epsilon}^i_L \,, \label{eq:q_sample}
\end{align}
where $\boldsymbol{\epsilon}_L$ and $\boldsymbol{\epsilon}_R$ are samples from a standard Normal distribution $\mathcal{N}(\mathbf{0},\mathbf{I})$, and $\Rmat$ and $\Lmat$ are defined in \cref{sec:cov_param} as part of our covariance parameterisation. 

As we are modelling a high-dimensional correlated distribution $\warpp \in \mathbb{R}^{3\times N_{v}}$, we might need to draw many samples to find effective ones, which would be computationally expensive and prohibitive in terms of memory to backpropagate through lots of low weighted samples. As such, we draw these samples without gradients but  store the noise components for use in a subsequent backwards pass through selected samples.

To identify the best samples to focus on, we compute our unnormalized importance weights without gradients, $\alpha^{i},$ based on the ratio between the target posterior and the proposal density
\begin{align}
    \alpha^{i} = \frac{p(\Ifix | \Imov \circ \warpp^{i}) p(\warpp^{i})}{q(\warpp^{i})} \,.
\end{align}
The normalized weights are then defined as $w^{i} = \alpha^{i} / \sum_{j=1}^{N_{s}} \alpha^{j}$. These weights represent the relative ``quality'' of each registration hypothesis under the image evidence.

To compute gradients for the network parameters, we form a weighted estimate of the posterior. Rather than backpropagating through all $N_{s}$ samples, many of which may have negligible weights, we perform multinomial resampling of these samples based on the normalised weight vector $\textbf{w}$. We draw $N_k$ indices, and then use these selected sample random noise vectors ($\boldsymbol{\epsilon}^k_R$ and $\boldsymbol{\epsilon}^k_L$) to redraw the samples and compute $\alpha^k$, but this time enabling gradients to flow back through the network to update the model weights. As we have now selected samples based on importance weighting, our final loss is simply the expected negative log-likelihood over these samples, that is
\begin{align}
    \mathcal{L}_{\mathrm{SIR}} = -\frac{1}{N_k} \sum_{k=1}^{N_k} \left( \log p(\Ifix | \Imov \circ \warpp^k) + \log p(\warpp^k) \right) \,.
\end{align}

\subsubsection{Likelihood Function}
Similarly to \citep{balakrishnan2019voxelmorph} we use normalised cross-correlation as our likelihood model, which provides robustness against global intensity fluctuations common in MRI acquisition. We treat it as the energy function of a Boltzmann distribution:
\begin{equation}
    p(\Ifix|\Imov, \warpp) \propto \exp\left( \frac{-\mathrm{NCC}(\Ifix, \Imov \circ \warpp))}{\sigma^2} \right) \,,
\end{equation}
where $\sigma^2$ is a likelihood scaling function.

\subsubsection{Prior Function}
For simplicity we use a diffusion based prior, where we penalise the squared spatial gradients of the deformation field:
\begin{align}
    p(\mathbf{Z}) \propto \exp \left( -\frac{\lambda}{2} \sum_{j=1}^{N_v} \| \nabla \mathbf{z}_j \|^2 \right) + \left( -\frac{\lambda_{\qmean}}{2} \sum_{j=1}^{N_v} \| \nabla \qmean_j \|^2 \right) \,.
\end{align}
We employ a stronger regularisation weight $\lambda_{\qmean}$ on $\qmean$ than on the sampled field $\warpp$ to allow the model to explore a wider range of the displacement manifold. This prevents the distribution becoming overly stiff, and enables the resulting samples to capture the multi-modal nature of the posterior. For fairness, we use the same prior in our variational experiments as well.

\subsection{Covariance Parameterisation}
\label{sec:cov_param}
We define the covariance matrix for our proposal, or variational, distribution as a sum of a low-rank and a sparse structured Cholesky factored matrix:
$\qcov = \Rmat \Rmat^\top + (\Lmat\Lmat^\top)^{-1}$ where $\Rmat \in \mathbb{R}^{N_v \times r}$ and $\Lmat \in \mathbb{R}^{N_v \times N_v}$ is a sparse lower triangular matrix with positive diagonal entries. 
In contrast to the 2D SUPN structure of \citep{dorta2018structured}, our sparsity structure for $\Lmat$ runs over the 3D spatial structure of the voxel grid and over the three channels of the output displacement vector.

Drawing samples from this is efficient, as we can draw from each covariance component independently, as noted in \cref{eq:q_sample}. We use the Cholespy solver~\citep{nicolet2021large} for efficient sampling using $\Lmat^{-\top}$.

To enable importance weighting, we need to evaluate the log likelihood under the proposal distribution of any sample $\warpp$, which is expressed as
\begin{equation}
    \log p(\warpp; \qmean, \qcov) = -\frac{1}{2} \bigg( \underbrace{(\vecf(\warpp)-\qmean)^\top \qcov^{-1} (\vecf(\warpp)-\qmean)}_{\qcov^{-1}\mathrm{\ quadratic\ prod}} + \underbrace{\log \det (2\pi \qcov)}_{\mathrm{logdet}} \bigg) \,,
\end{equation}
where the displacement field is vectorised, $\vecf(\warpp) \in \mathbb{R}^{3N_v}$.
The quadratic inverse covariance product can be evaluated, by applying the Woodbury Matrix Inversion Lemma, as
\begin{equation}
    (\vecf(\warpp)-\qmean)^\top \qcov^{-1} (\vecf(\warpp)-\qmean) = \textbf{k}^\top \textbf{k} - (\textbf{S}^\top \textbf{k})^\top \textbf{M}^{-1}(\textbf{S}^\top \textbf{k}) \,,
\end{equation}
where $\textbf{k} = \Lmat^\top (\warpp-\qmean)$, $\textbf{S}=\Lmat^\top \Rmat$ and $\textbf{M} = \textbf{I}^{-1} + \textbf{S}^\top\textbf{S}$, and $\textbf{M} \in \mathbb{R}^{r \times r}$, which we invert using an eigenvalue decomposition (to avoid numerical instabilities). The log determinant term can be evaluated as
\begin{equation}
\log \det \qcov = \log\det ((\Lmat\Lmat^\top)^{-1} + \Rmat\Rmat^\top) = -\log\det (\Lmat\Lmat^\top) + \log\det (\textbf{M}) \,.
\end{equation}
A full derivation is given in \cref{sec:deriv}.

\subsection{Network Architecture}
We use a fully convolutional U-net style architecture, without bias terms. The model has 4 resolution levels, with 16 channels for the first resolution level of the encoder, and 24 channels thereafter. We have separate convolutional prediction heads for the mean, $\qmean$, the sparse Cholesky factor $\Lmat$, and the low-rank matrices $\Rmat$. The first two are calculated at half the voxel resolution (to improve efficiency) and upsampled using a cubic spline, for $C^2$  smoothness; 
the low-rank matrices are produced at half that resolution level, to reduce memory demands, and upsampled to match the Cholesky factor using bilinear interpolation. No activation functions are applied to $\qmean$ or $\Rmat$; for $\Lmat$ we use a soft-plus for the Cholesky diagonals, to ensure they are positive, and no activation for the off-diagonal terms.

\section{Experiments}
\subsection{Implementation Details}
All models were trained for 50,000 steps with a batch size of 1 on a single NVIDIA RTX A6000 (48 GB VRAM), with a peak memory requirement of 32 GB. We utilized the SOAP optimizer \citep{vyas2024soap} via the Heavyball library \citep{heavyball}, which facilitates stable training at high learning rates; we set the learning rate to $ 2\times 10^{-3}$ with $\beta_1 = 0.9$ and $\beta_2 = 0.95$. A cosine learning rate scheduler was employed, decaying to $2\times 10^{-4}$ after 30,000 iterations. To improve model robustness, we applied data augmentation consisting of coherent voxel shifts (up to 8 voxels) and independent (to each image in the pair) voxel shifts (up to 2 voxels).

\subsection{Hyperparameters}
We fix the number of low-rank components to 25, unless noted otherwise and provide additional results in \cref{sec:app_num_low_rank}. The Cholesky connectivity pattern $\Lmat$ accounts for neighbors within a $3\times 3\times 3$ kernel with cross-channel correlations between deformation directions, larger neighbours would require too much VRAM. We apply $L_2$ regularization with a coefficient of $0.05$ to the off-diagonal elements of $\Lmat$ to prevent overfitting and ensure numerical stability during inference.

We use a ratio of $\lambda_{\qmean} = 2.5, \lambda = 1.0$ and tuned the likelihood scale parameter to $\sigma = 0.5$ on the variational baseline for segmentation accuracy while maintaining less than 0.05\% folding of the deformation field. 

For the SIR-based variants, we introduce a temperature scaling hyperparameter to prevent the importance weight distribution becoming too spiky (dominated by individual samples), or flat (uniform sampling). We adopt a dynamic temperature scaling approach\citep{neal2001annealed}, where the unnormalised weights $\boldsymbol{\alpha}$ are scaled by a factor of $\frac{T}{\sigma_{\alpha}}$. Here $\sigma_{\alpha}$ is an exponential moving average of standard deviation of the weights, i.e. $\sigma^{t}_\alpha = \gamma \, \sigma^{t-1}_\alpha + (1-\gamma) \, \mathrm{std}(\boldsymbol{\alpha)}$, where $t$ indicates the optimisation step and $\gamma = 0.9$. We select $T=3.0$, which empirically provides a trade-off between uniform, and highly selective weights. Further temperature experiments are in \cref{sec:app_temperature}.

\subsection{Inference and Posterior Characterisation}
At inference  time, we draw $N_l=1,200$ samples (in batches), and perform importance weighting on these. Through multinomial sampling, we then draw a total of $N_k=80$ samples to characterise the posterior distribution.
While we use the final EMA value of $\sigma_\alpha$ obtained from training for test-time scaling, we note that this could be optimised post-training to improve performance.

\subsection{Model Variants}
We evaluate the performance of our model in terms of propagated segmentation accuracy after registering pairs of images. We denote the baseline variational variants of our model with a leading \textbf{V}, regular importance sampling has a leading \textbf{I}, and SIR with \textbf{S}. Symbols after the + represent the covariance parameterisation: \textbf{D} is a diagonal Cholesky, \textbf{C} is a sparsely structured Cholesky and \textbf{L} indicates a low-rank component. We also compare against a resolution matched Voxelmorph model\citep{dalca2019unsupervised}, with a comparable architecture (see \cref{sec:app_baselines} for details) that we ensembled using 5 models (Vxm5).

\subsection{Data}
We use a subset of the OASIS \citep{marcus2007open} as pre-processed and described in \citep{balakrishnan2019voxelmorph}. Each image has 35 automatically segmented labels using \citep{fischl2012freesurfer}, 17 of which are bilateral, and is resampled to $1\ mm^3$ isotropic voxels with an image size of $160 \times 192 \times 224$. The resolution of the predicted grid, following cropping some of the background, is $96 \times 72 \times 80$, leading to $N_v = 552,960$ and considering a 3D displacements field leads to us estimating a multivariate distribution over $1.6$ million dimensions.

The data contains substantial anatomical variations and differences in brain shape. It also has dense high-resolution anatomical labels of varying scales and morphological complexity, enabling us to validate the uncertainty estimates that arise. During both training and evaluation, we perform subject-to-subject registration, where each image is registered to others within its respective set. For the test set, this results in $50 \times 49 = 2,450$ unique registration pairs to ensure a rigorous assessment of model generalisation and uncertainty calibration.

\section{Results}
\subsection{Accuracy}
\begin{figure}
    \centering
    \includegraphics[width=0.48\linewidth]{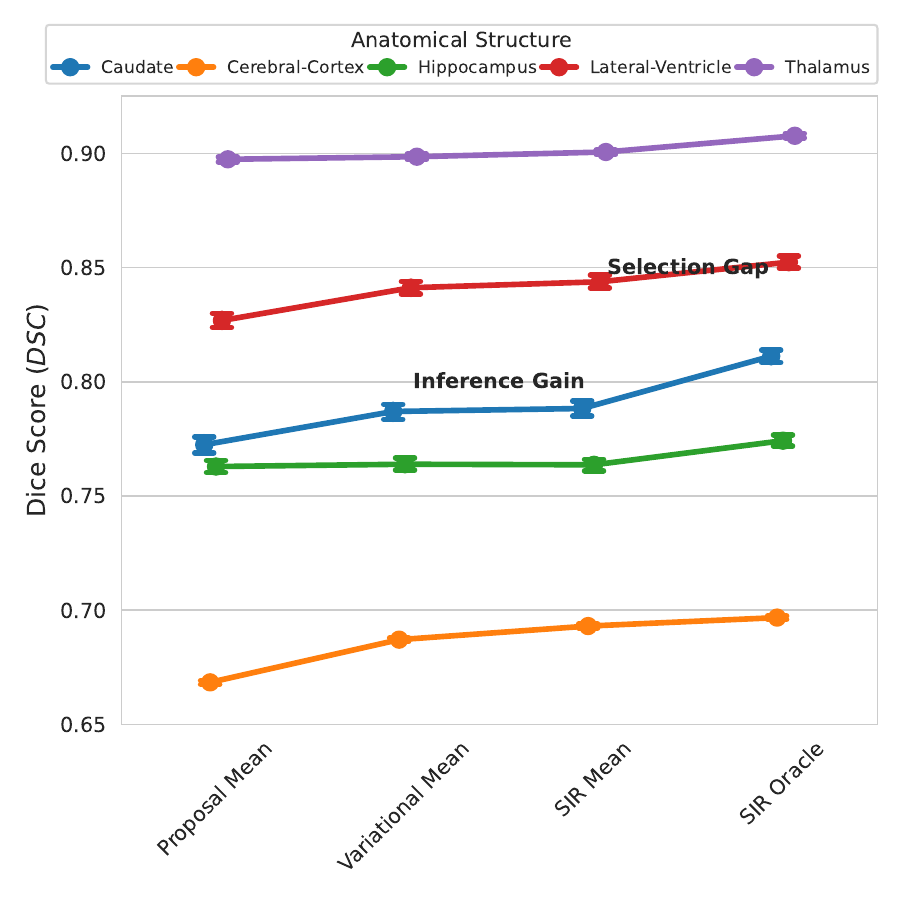}
    \includegraphics[width=0.48\linewidth]{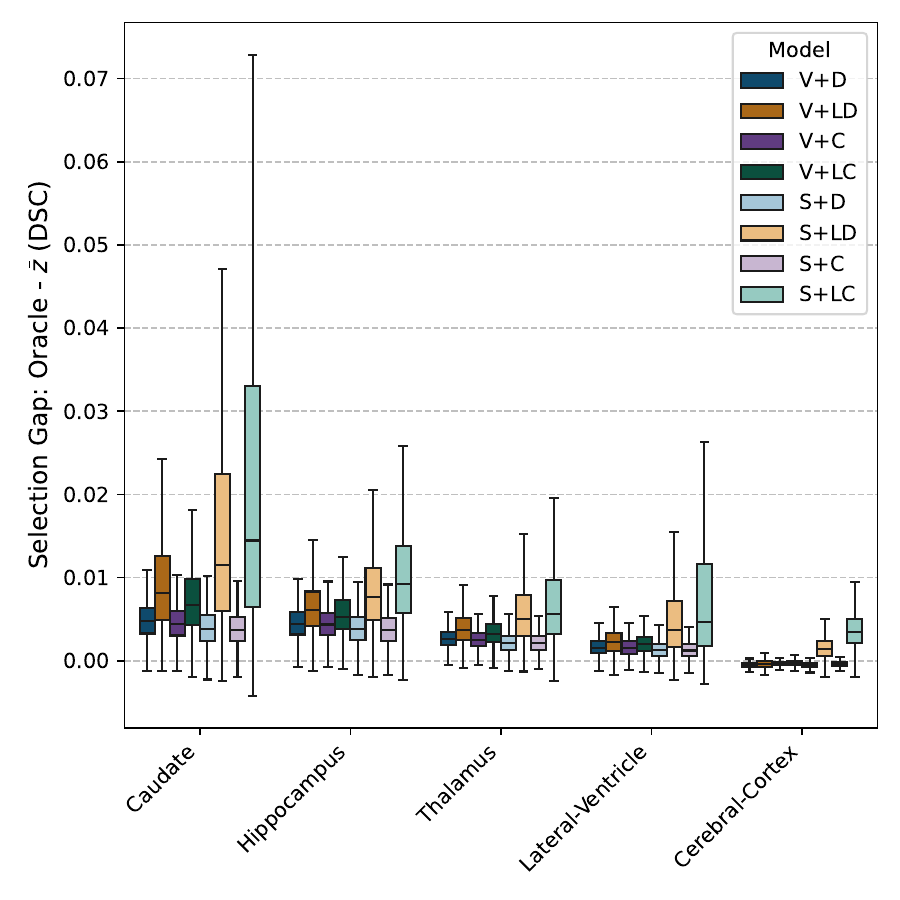}
    \caption{Left, an illustration of the proposal mean ($\qmean$ in SIR), variational $\qmean$, SIR sample mean and Oracle (best-sample), and standard deviation of the segmentation accuracy for different model variants. Right, a boxplot exemplifying the quality of the Oracle sample drawn from the approximate posterior distribution. Larger Dice Similarity Coefficient (DSC) values indicated greater segmentation accuracy.}
    \label{fig:ladder}
\end{figure}

\begin{table}
\centering
\caption{Registration performance and folding rates across OASIS test pairs. We report the Dice Similarity Coefficient (DSC) for the mean prediction ($\mu_{\warpp}$), the mean of resampled fields ($\bar{\warpp}$), and the Oracle (the single sample from the posterior with the highest DSC). Fold \% represents the percentage of voxels with a non-positive Jacobian determinant. Standard deviations are shown in parentheses. High DSC values and lower fold \% are preferable.}
\label{tab:model_comparison}
\begin{tabular}{l|ccc|ccc}
\toprule
 & $\qmean$ (DSC) & $\bar{\warpp}$ (DSC) & Oracle (DSC) & $\qmean$ fold \% & $\bar{\warpp}$ fold \% & Oracle fold \% \\
Model &  &  &  &  &  &  \\
\midrule
V+D & 0.800 (0.04) & 0.800 (0.04) & 0.805 (0.04) & 0.046 (0.04) & 0.059 (0.05) & 0.059 (0.05) \\
V+LD & 0.799 (0.04) & 0.798 (0.04) & 0.805 (0.04) & 0.046 (0.04) & 0.059 (0.05) & 0.059 (0.05) \\
V+C & 0.802 (0.04) & 0.802 (0.04) & 0.807 (0.04) & 0.045 (0.05) & 0.057 (0.06) & 0.057 (0.06) \\
V+LC & 0.800 (0.04) & 0.800 (0.04) & 0.806 (0.04) & 0.044 (0.03) & 0.056 (0.04) & 0.056 (0.04) \\
\midrule
I+LC & 0.800 (0.04) & 0.800 (0.04) & 0.805 (0.04) & 0.050 (0.05) & 0.063 (0.06) & 0.063 (0.06) \\
\midrule
S+D & 0.801 (0.04) & 0.801 (0.04) & 0.805 (0.04) & 0.048 (0.05) & 0.062 (0.06) & 0.062 (0.06) \\
S+LD & 0.798 (0.04) & 0.800 (0.04) & 0.809 (0.04) & 0.032 (0.03) & 0.085 (0.08) & 0.096 (0.10) \\
S+C & 0.801 (0.04) & 0.801 (0.04) & 0.805 (0.04) & 0.051 (0.03) & 0.065 (0.04) & 0.065 (0.04) \\
S+LC & 0.794 (0.04) & 0.801 (0.04) & 0.812 (0.03) & 0.011 (0.01) & 0.115 (0.07) & 0.163 (0.13) \\
\midrule
S+LC & 0.788 (0.04) & 0.802 (0.04) & 0.814 (0.03) & 0.004 (0.01) & 0.240 (0.14) & 0.292 (0.25) \\

10L &  &  &  &  &  & \\
\midrule
Vxm5 & 0.798 (0.03) & 0.798 (0.03) & 0.804 (0.03) & 0.000 (0.00) & 0.000 (0.00) & 0.000 (0.00) \\
\bottomrule
\end{tabular}
\label{tab:accuracy}
\end{table}

\cref{fig:ladder} illustrates the improvements in the Dice Similarity Coefficient (DSC) for specific structures when using S+LC over V+LC. Most strikingly, we observe a substantial selection gap, which we define as the improvement in the Oracle (most accurate) segmentation performance, compared to the SIR or Variational mean. This is further examined in \cref{tab:accuracy}, where we summarise the average segmentation propagation accuracies and observe all approaches have similar average performance but S+LC and S+LD show the biggest gains. A complete set of boxplots for different structures is provided in \cref{sec:accuracy_app}.  
The more complex posterior forms offer limited benefits in variational inference but confer a pronounced effect in SIR models. In our sensitivity analysis (\cref{sec:app_num_low_rank}), we found that 10 low-rank components outperforms in 25 in terms of accuracy, and calibration, while also being more efficient to sample.
We also observe a slight improvement in accuracy over the 5 model Voxelmorph ensemble, at the expense of folding.
We note that the level of voxel folding increases with the more exploratory, high-variance deformations found by SIR, which are particularly noticeable in the Oracle samples. Further evaluation of the oracle samples is given in \cref{sec:selection_app}.

\subsection{Calibration of the Posterior}

\begin{figure}[t]
    \centering
    \includegraphics[width=\linewidth]{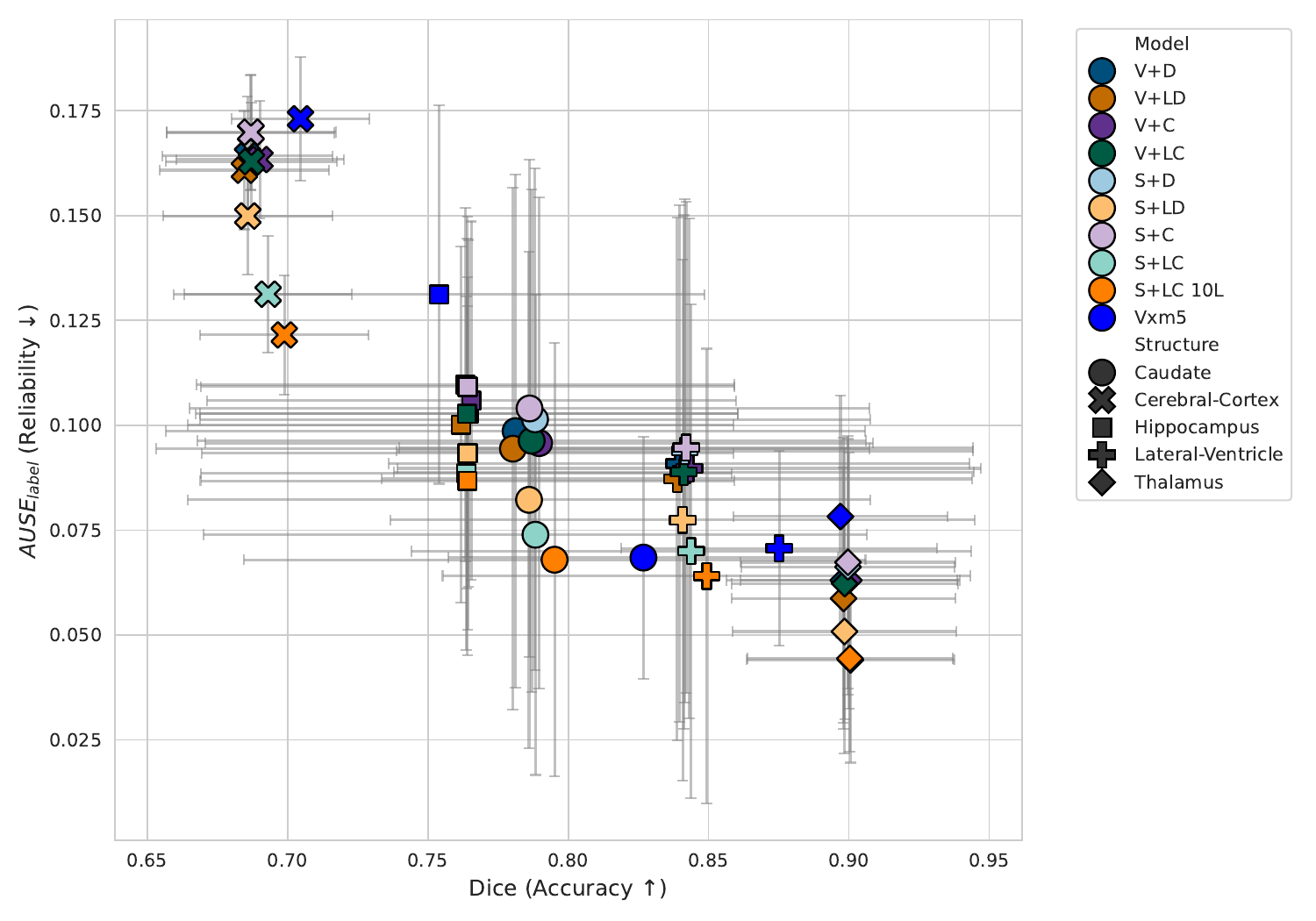}
    \caption{Accuracy vs calibration for the different model variants for a set of structures, standard deviations plotted in gray. The best performance is in the bottom-right of the graph with a low AUSE (better calibration) and a high DSC (better accuracy). The full version of the model (S+LC) with 10 low-rank components offers improved calibration with equal or improved accuracy consistently over all structures. The 5 model Voxelmorph ensemble has improved average accuracy for some structures, but generally much worse calibration.}
    \label{fig:accuracy_vs_calibration}
\end{figure}

\begin{table}
\centering
\caption{Uncertainty Calibration Metrics: Area Under Sparse Error (AUSE) based on label entropy and Expected Calibration Error (ECE) calculated using a dilated (by 3 voxels) segmentation mask. The last column is the Spearman correlation of label entropy and Dice. All fields show the average and std-dev over structures.}
\label{tab:calibration_metrics}
\begin{tabular}{l|p{22mm} p{22mm} p{22mm} p{27mm}}
\toprule
 Model & $\mathrm{AUSE}_{\mathrm{label}} \downarrow$ & $\mathrm{ECE} \downarrow$ & Label Entropy & Mean Spearman r: Entropy \& DSC \\
\midrule
V+D & 0.0850 (0.021) & 0.1285 (0.027) & 0.0628 (0.002) & 0.064 (0.169) \\
V+LD & 0.0818 (0.022) & 0.1257 (0.027) & 0.0705 (0.002) & 0.085 (0.165) \\
V+C & 0.0854 (0.021) & 0.1279 (0.026) & 0.0595 (0.002) & 0.046 (0.164) \\
V+LC & 0.0844 (0.021) & 0.1274 (0.027) & 0.0641 (0.002) & 0.033 (0.167) \\
\midrule
I+LC & 0.0861 (0.021) & 0.1287 (0.027) & 0.0621 (0.002) & 0.011 (0.172) \\
\midrule
S+D & 0.0897 (0.021) & 0.1312 (0.026) & 0.0549 (0.007) & 0.051 (0.123) \\
S+LD & 0.0736 (0.020) & 0.1159 (0.026) & 0.0901 (0.004) & -0.137 (0.177) \\
S+C & 0.0907 (0.021) & 0.1324 (0.027) & 0.0531 (0.006) & 0.007 (0.121) \\
S+LC & 0.0659 (0.019) & 0.1064 (0.025) & 0.1079 (0.006) & -0.325 (0.218) \\
\midrule
S+LC 10L & 0.0636 (0.018) & 0.1028 (0.023) & 0.1156 (0.010) & -0.419 (0.217) \\
\midrule
Vxm5& 0.1008 (0.015) & 0.1224 (0.020) & 0.0703 (0.007) & -0.410 (0.209) \\
\bottomrule 
\end{tabular}
\label{tab:calibration}
\end{table}

We evaluate calibration using both the Expected Calibration Error (ECE) and Area Under the Sparsification Error (AUSE) for each segmentation region. 
To focus our metric on relevant regions, we compute ECE and AUSE within a 3-voxel dilated boundary of the anatomical labels, as registration ambiguity and label uncertainty are most pronounced at tissue interfaces, specific details given in \cref{app:uncertainty_metrics}. ECE quantifies the average discrepancy between a model's predicted probability of a label and the actual observed frequency of that label; in the context of registration, a lower ECE indicates that the model's confidence scores are more reliable indicators of anatomical alignment. Whereas, AUSE evaluates uncertainty by measuring how registration error decreases as we progressively remove voxels with the highest uncertainty; a lower AUSE signifies that the uncertainty map is an effective proxy for localized registration errors

We evaluate the calibration of our models in \cref{tab:calibration}. We find that our SIR models with low-rank covariance parameterisations, particularly S+LC with 25 or 10 low-rank components, are much better calibrated than the variational models, the naive importance sampling approach (which performs similarly to variational inference), and the Voxelmorph deep ensemble in terms of both ECE and AUSE. We provide a structure-wise evaluation in \cref{sec:calibration_app}.
Furthermore, we examined the correlation between label entropy and accuracy (in terms of DSC) for each structure, and averaged these. We find that the poorly calibrated models exhibit small \textbf{positive} correlations, incorrectly indicating more uncertainty when they are most accurate. Whereas, our better calibrated models, such as S+LC and S+LC 10L, demonstrate strong negative correlations indicating the uncertainty is a much more reliable proxy for registration accuracy. 

We highlight the relationship between accuracy and calibration for the different model variants in Fig \ref{fig:accuracy_vs_calibration}, which clearly illustrates that S+LC leads to improved calibration without any  implications on accuracy. 

\subsection{Multi-Modality}

\begin{figure}
    \centering
    \includegraphics[width=\linewidth]{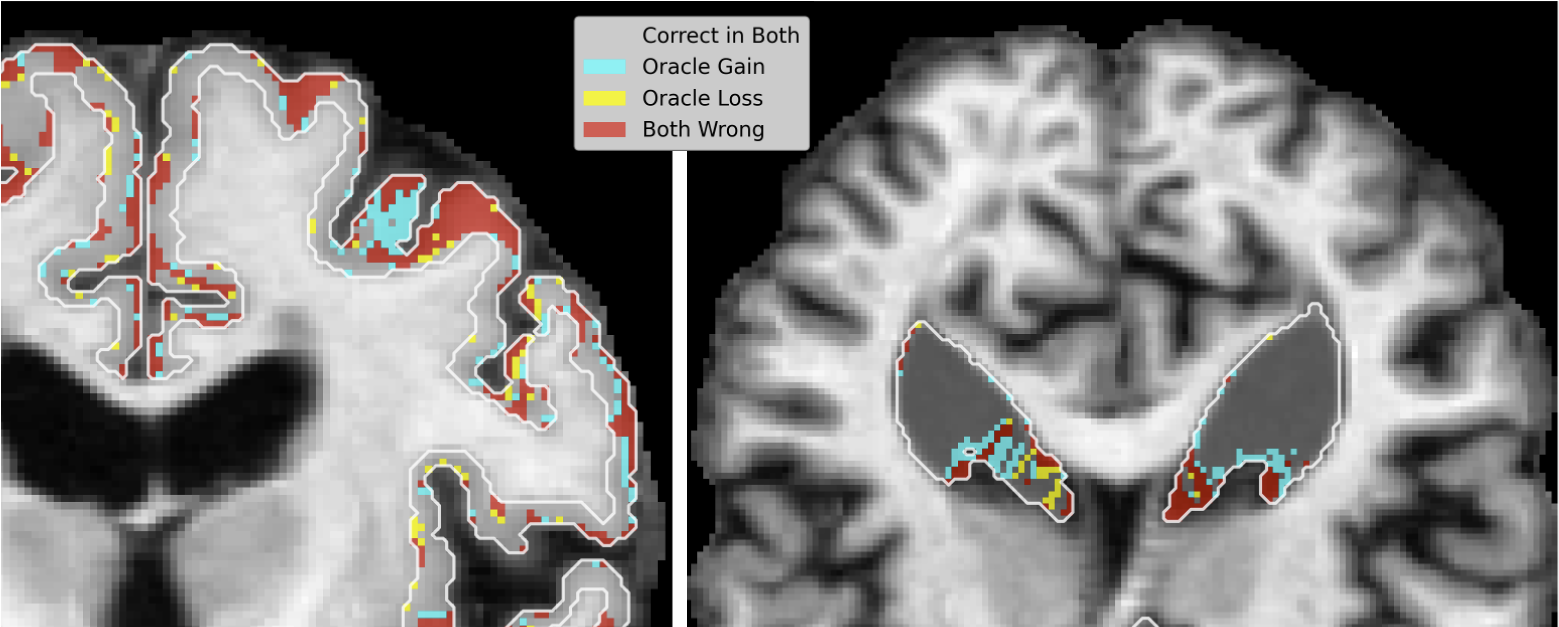}
    \caption{Illustration of the structured nature of the samples, where we find the performance of the Oracle sample can lead to substantial structured gains over the average SIR deformation. On the left, we show an example where a lobe of the cerebral cortex is correctly aligned by the oracle sample, with a DSC improvement of 0.04. On the right, we observe substrantial structured changes in the lateral ventricle, with a DSC improvements of 0.03. }
    \label{fig:oracle_samples}
\end{figure}

The spatial coherence, and multi-modality, of the learned distribution is exemplified by examining the improvement in segmentation performance by the Oracle sample. As shown in \cref{fig:oracle_samples}, we can see segmentation corrections in smooth groups of pixels, which is enabled through our learned spatial distribution.

\begin{figure}
    \centering
    \includegraphics[width=0.45\linewidth]{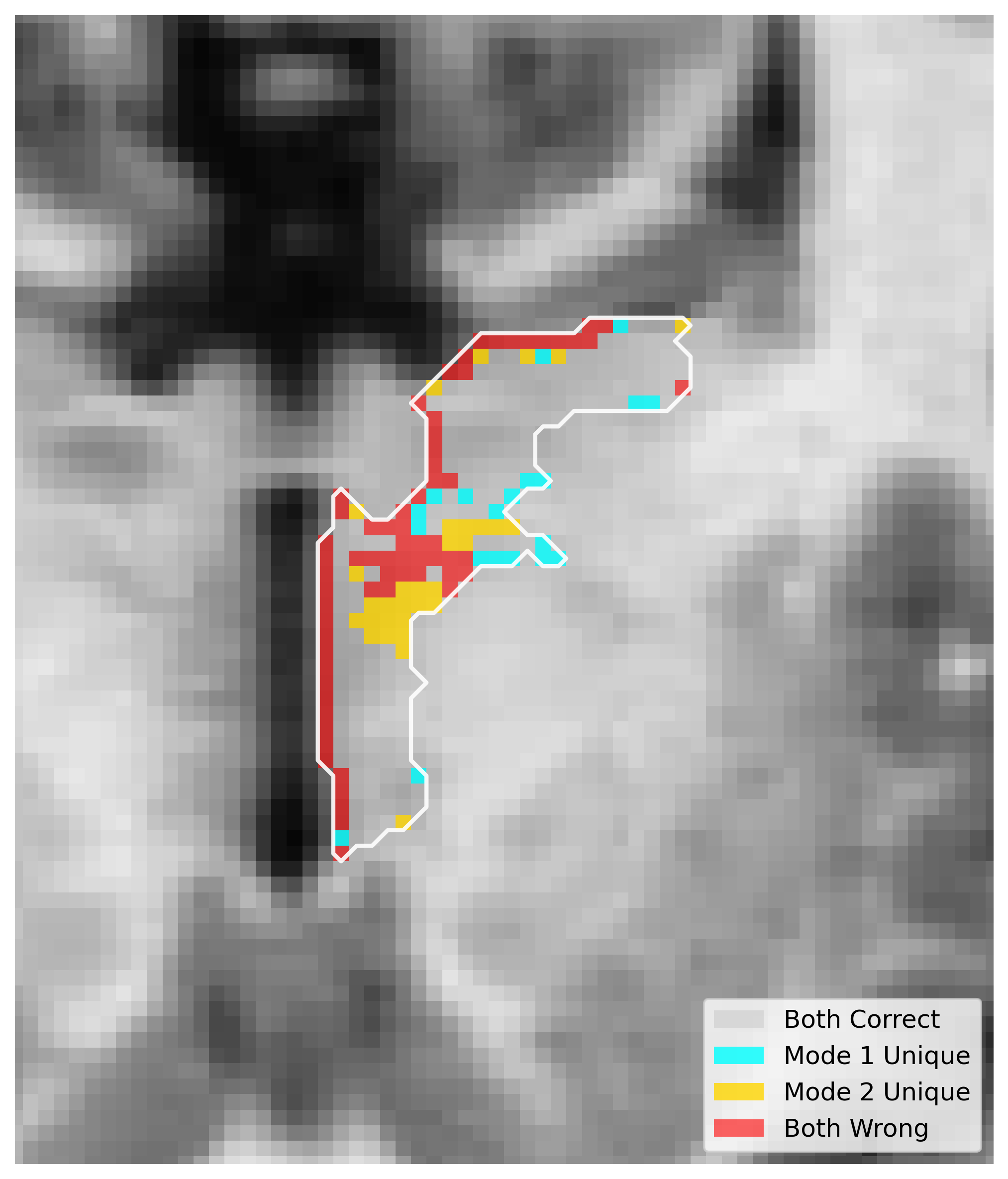}
    \includegraphics[width=0.45\linewidth]{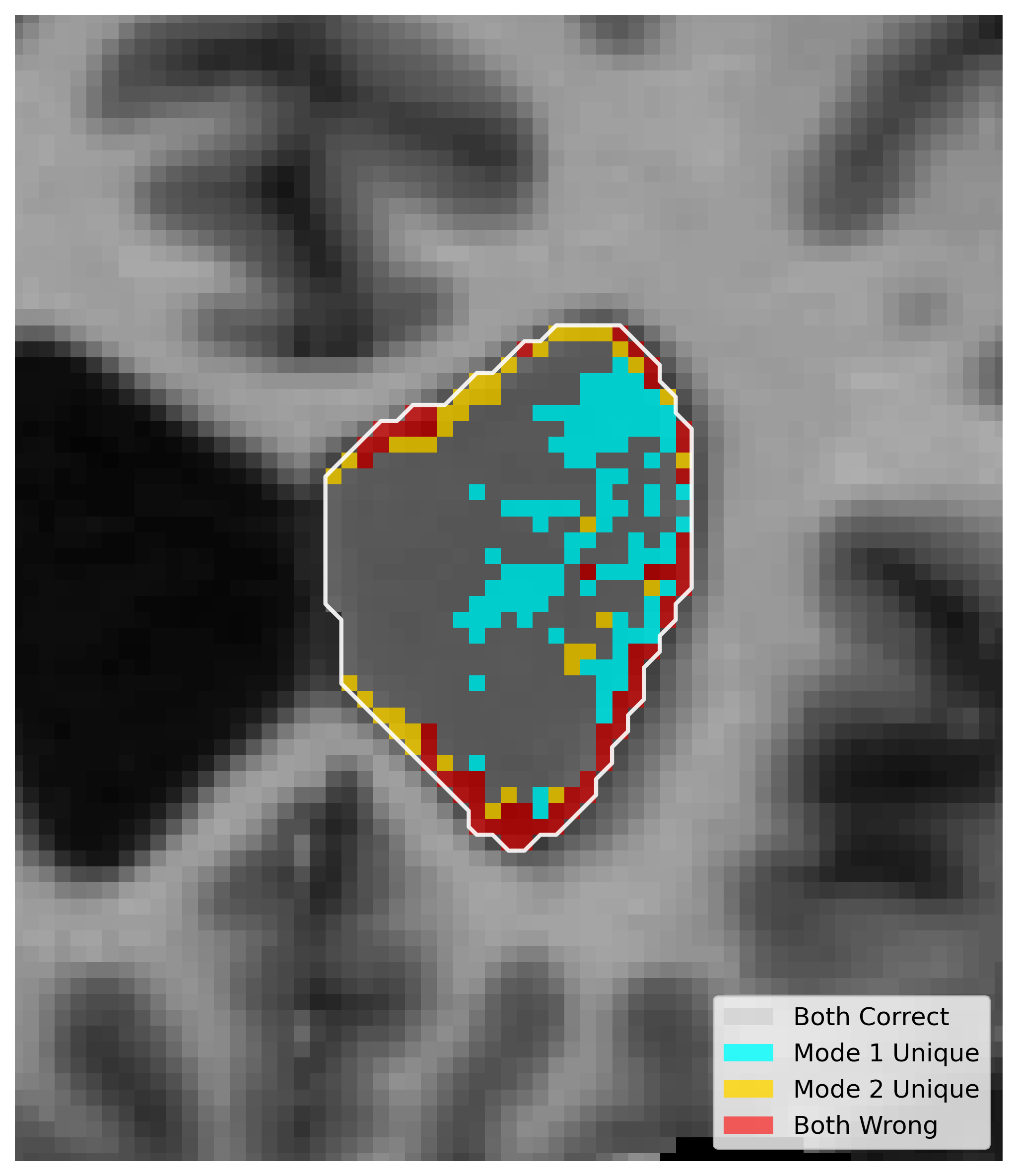}
    \caption{Illustration of the multi-modality of the posterior distribution, where disagreements between modes are marked in blue or orange. On the left, we find structured changes in the Thalamus where mode 1 has a DSC of 0.834 and mode 2 has a Dice of 0.824. On the right, we observe substantial differences in the left lateral ventricle, either on the interior or exterior border, with a DSC of 0.864 for mode 1 and 0.862 for mode 2.}
    \label{fig:multimodality}
\end{figure}
To inspect individual cases, we performed PCA on the SIR sample deformations, run clustering to identify groups (in our case the top two) and then picked examples nearest to those cluster means. Two examples of bimodality with equivalent DSC accuracy are given in~\cref{fig:multimodality}.

\section{Discussion}
Our results demonstrate that our proposed SIR approach, when coupled with structured covariance parameterisations, significantly improves the calibration of registration uncertainty without compromising accuracy.  By moving beyond the restrictive assumptions of amortised variational inference, we demonstrate that registration uncertainty can be characterised in a manner that serves as a reliable proxy for anatomical alignment.

The SIR framework acts as a consistent surrogate to the importance weighted objective in \cref{eq:iwae}, analogous to the Reweighted Wake-Sleep 'wake' update~\citep{bornschein2014reweighted}, stabilizing training by decoupling sample weighting from gradient propagation. While not a theoretically guaranteed optimizer of \cref{eq:iwae}, our empirical results and their stability over training \cref{sec:stability}, demonstrate that this strategy effectively guides the proposal distribution toward high-probability regions while avoiding the entropy collapse seen in direct importance weighted optimisation.

\subsection{Calibration and Meaningful Uncertainty}
A central finding in this work relates to the relationship between model uncertainty and segmentation error. Our proposed S+LC model significantly outperforms variational baselines and Voxelmorph ensembles in terms of calibration metrics, but also exhibits a strong negative correlation between label entropy and accuracy. 
The expanded entropy range of the proposed model, coupled with its superior registration accuracy, suggests that the captured uncertainty is highly structured within the high-dimensional deformation space, representing a coherent distribution of anatomically plausible hypotheses rather than stochastic noise. 

\subsection{The Accuracy of Oracle Samples}
The selection gap between the SIR mean, and the Oracle provides an intuition into the complexity of the inferred posterior. It indicates that very good samples exist with high-probability in the proposal distribution, but perhaps due to multi-modality do not coincide with the mean. This finding supports the use of SIR for fitting complex posterior distributions.

\subsection{Computational Efficiency}
Despite the high dimension of the displacement field, our approach remains computationally efficient, requiring a single forward pass of the amortised model to produce an effective proposal distribution from which samples are drawn. Samples require a sparse linear solve for $\mathbf{L}$, which is $O(\text{nnz}(\mathbf{L}))$, and a matrix multiplication for $\mathbf{R}$. The entire process of proposal inference, sample drawing and weighting of 1,200 samples is completed in under 5 seconds on a single GPU. 

This represents a substantial computational benefit over hierarchical sampling approaches such as PULPo~\citep{siegert2024pulpo}, which require sequential forward passes through the model for each sample, coupling the sampling complexity with the depth of the neural network, or MCMC-based methods \citep{grzech2021uncertainty}, which require multiple iterative passes.

\subsection{Limitations and Future Directions}
While our approach improves calibration, with good accuracy, there is a trade-off in terms of topological regularity and folding of voxels. Future work could investigate using stationary velocity fields, as in \citep{dalca2019unsupervised}, to encourage diffeomorphic transformations. Furthermore, more complex regularisers \citep{burger2013hyperelastic} may offer benefits in certain applications.

Our architecture was chosen based on our existing code library, however the underlying SIR and covariance parameterisation are model-agnostic. Future directions could replace the convolutional backbone with a vision transformer architecture \citep{chen2022transmorph}.

In terms of improvement to segmentation calibration, we could investigate post-training optimisation of the temperature parameter, and more rigorous hyper-parameter selection. We could also include segmentation losses within the likelihood function; although these are inaccessible at test time, but they could contribute to learning the shape of the importance sampling distribution.

Beyond medical image registration, our proposed integration of SIR with our structured sparsity and low-rank covariance model offers a versatile framework for high-dimensional spatial inference tasks in inverse problems that suffer from ambiguity, such as optical flow~\citep{ilg2018uncertainty}, sparse-view CT or 3D image reconstruction and image super-resolution.

\section{Conclusion}
We have introduced Structured SIR, a framework for efficient and expressive approximate inference of posterior distributions in high-dimensional imaging tasks. We have demonstrated the application of this to probabilistic image registration, where we leverage SIR with a novel structured covariance parameterisation to characterise a 1.6 million dimensional posterior distribution of a 3D deformation field. 

Our proposed SIR algorithm makes use of a dual sampling pass: first evaluating a large set of candidates to weight based on their respective probabilities, then performing resampling of these to characterise the posterior. To the best of our knowledge, this is the first application of SIR to a dense vision task of this scale.

Our results demonstrate that SIR with sufficiently complex covariance parameterisations, exhibits good calibration performance, substantially outperforming variational inference while maintaining accuracy. Moreover, the substantial gap in the accuracy of the Oracle sample, confirms that our expressive structured proposal distribution supports a diverse range of plausible deformations. 

This frameworks offers a scalable blueprint for uncertainty quantification in high-dimensional spatial tasks. While demonstrated here for registration, it could be extensible to other inverse problems where test-time evidence can guide the selection of solutions.

\section*{Acknowledgements}
NC acknowledges support from the EPSRC CAMERA Research Centre \\(EP/T022523/1), the UKRI Strength in Places Fund My-World Project \\(SIPF00006/1) and the Royal Society.
We thank Teo Deveney and Paula Seidler for their help refactoring the supporting codebase for Cholesky calculation and sampling.
\bibliographystyle{plainnat}
\bibliography{refs}
\appendix
\newpage

\section{Covariance Formulation}
\label{sec:deriv}

We define our covariance matrix as
\begin{equation}
    \qcov = \Rmat\Rmat^\top + (\Lmat\Lmat^\top)^{-1} \,,
\end{equation}
where $\Rmat \in \mathbb{R}^{n\times r}$ and $\Lmat \in \mathbb{R}^{n\times n}$ (but sparse).
Samples are easy to obtain, as per \cref{eq:q_sample}, as a draw from each covariance component.

We need to consider the log probability under this density
\begin{equation}
    \log p\big(\vecwarpp; \qmean, \qcov\big) = -\frac{1}{2} \left( \underbrace{\xb^\top \qcov^{-1} \xb}_{\qcov^{-1}\mathrm{ prod}} + \underbrace{\log \det (2\pi \qcov)}_{\mathrm{logdet}} \right) \,,
\end{equation}
where we have $\xb := \vecwarpp - \qmean$ for generalisation purposes, and to aid readability.
We derive a computational tractable form for each term individually in the following sections.

\subsection{Inverse Covariance Product}

Considering
\begin{align}
    \xb^\top \qcov^{-1} \xb = \xb^\top \left(\Rmat\Rmat^\top + (\Lmat \Lmat^\top)^{-1}\right)^{-1} \xb \,,
\end{align}
we can apply the Woodbury Matrix identity
\begin{align}
    (\A + \U\C\V)^{-1} = \A^{-1} - \A^{-1}\U(\C^{-1} + \V\A^{-1}\U)^{-1}\V\A^{-1} \,,
\end{align}
where
$\A = (\Lmat \Lmat^\top)^{-1}$, so $\A^{-1} = \Lmat\Lmat^\top$,
$\V = \U^\top = \Rmat^\top$ and $\C=\mathbf{I}$.
Therefore, we seek to calculate
\begin{align} \xb^\top\qcov^{-1}\xb &= \xb^\top (\A + \U\C\V)^{-1} \xb \\
&=   \xb^\top \A^{-1}\xb - \xb^\top \A^{-1}\U(\C^{-1} + \V\A^{-1}\U)^{-1}\V\A^{-1}\xb \,,
\end{align}
where $\kb := \Lmat^\top \xb$, which we can efficiently calculate via convolution operations~\citep{dorta2018structured}, reducing the first term of the expression to a simple quadratic product $\kb^\top \kb$.

Turning our attention to the second term, if we denote
\begin{equation}
\M = \C^{-1} + \Rmat^\top \A^{-1}\Rmat \,,
\end{equation} 
where $\M \in \mathbb{R}^{r \times r}$ and note that
\begin{equation}
    \V\A^{-1}\xb = \V\Lmat\Lmat^\top \xb = \V\Lmat \kb = \Sb^\top \kb \,,
\end{equation}
where $\Sb^\top := \V\Lmat$. Therefore $\Sb=\Lmat^\top \U$ that we can again calculate efficiently through convolving $\Lmat^\top$ with the low-rank images in $\Rmat$. We can then write $\M = \C^{-1} + \Sb^\top \Sb$.

Putting it all together we have
\begin{align} \xb^\top\qcov^{-1}\xb 
&= \xb^\top (\A + \U\C\V)^{-1} \xb \\
&=   \xb^\top \A^{-1}\xb - \xb^\top \A^{-1}\U(\C^{-1} + \V\A^{-1}\U)^{-1}\V\A^{-1}\xb \\
&= \kb^\top \kb - \xb^\top \Lmat \Lmat^\top \U \M^{-1}\V \Lmat \Lmat^\top \xb \\ 
&= \kb^\top \kb - (\kb^\top \Sb)^\top \M^{-1}(\Sb^\top \kb) \\ 
&= \kb^\top \kb- (\Sb^\top \kb)^\top \M^{-1}(\Sb^\top \kb) \,.
\end{align}
As $\M$ is only of size $r \times r$, where $r << n$ (in this paper $r=25$), calculation of a Cholesky factorisation, or Eigen-decomposition, is computationally tractable. In practice, we use an Eigen-decomposition to avoid numerical issues that can arise with very small Eigenvalues.

\subsection{Log-Determinant}
Finally, we can apply the matrix determinant lemma using the definition of $\M$ above to give
\begin{equation}
\log \det \qcov = \log\det \big((\Lmat \Lmat^\top)^{-1} + \Rmat\Rmat^\top \big) = -\log\det (\Lmat\Lmat^\top) + \log\det (\M) \,,
\end{equation}
where $\log\det (\M)$ falls out of the Eigenvalue decomposition required for the log probability above.

\newpage
\section{Segmentation}
\label{sec:accuracy_app}
\subsection{Label-wise Accuracy}
\begin{figure}[ht!]
    \centering
    \includegraphics[width=0.9\linewidth]{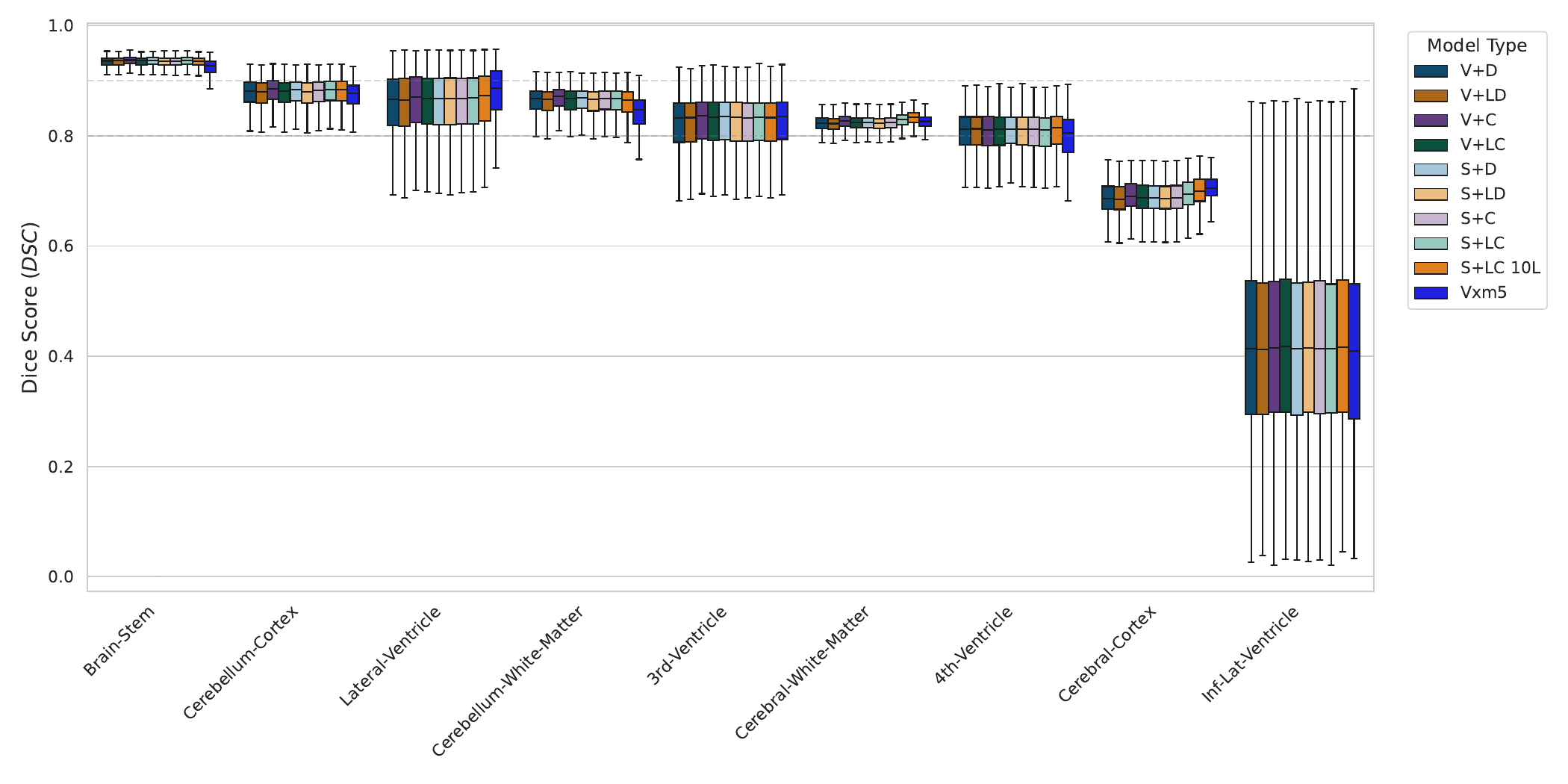}
    \includegraphics[width=0.9\linewidth]{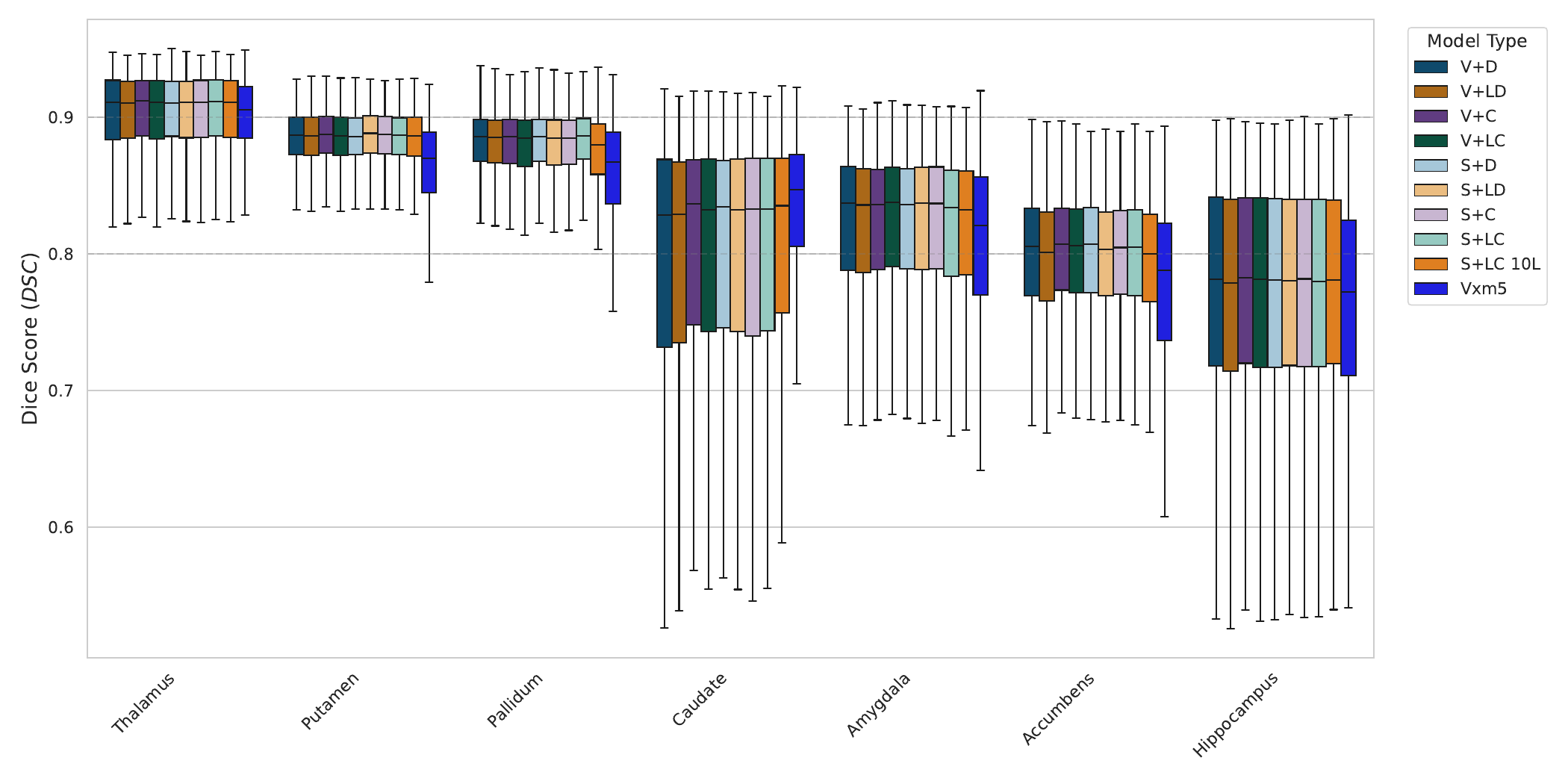}
    \caption{Boxplots of propagated segmentation accuracy for different labelled region for different model variants, when using the average of sampled fields $\bar{\warpp}$.}
\end{figure}

\FloatBarrier\newpage

\section{Selection Gap}
\label{sec:selection_app}
\begin{figure}[b!]
    \centering
    \includegraphics[width=0.65\linewidth]{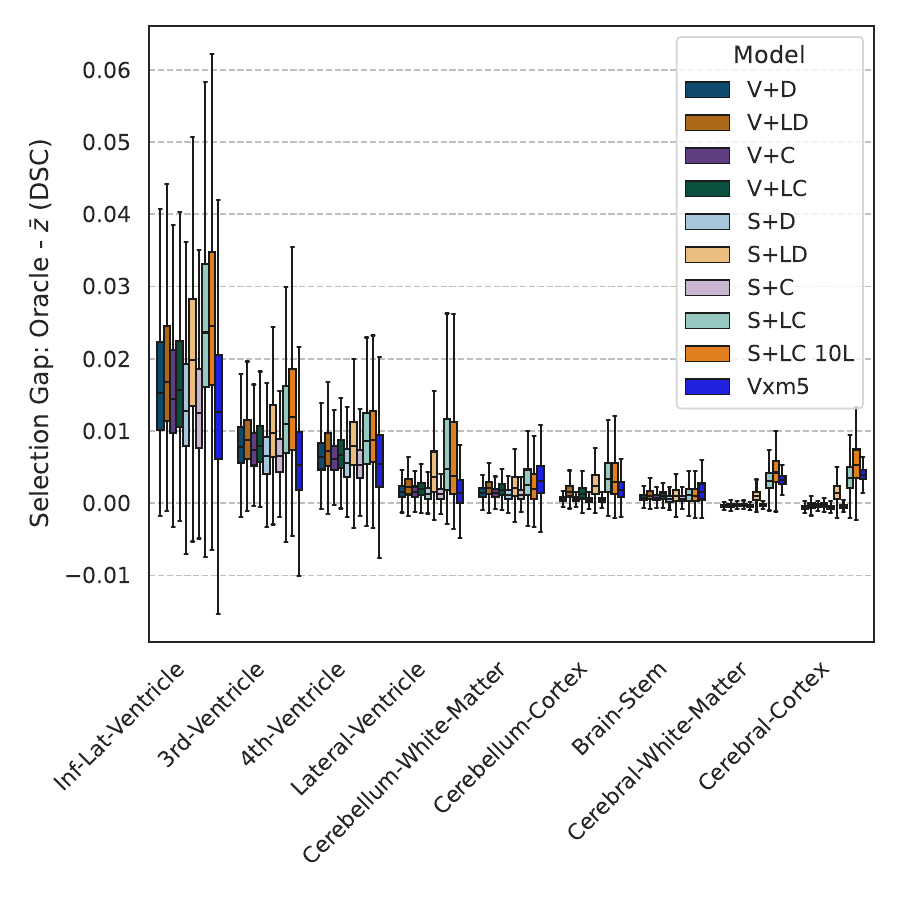}\\
    \vspace{-0.3cm}
    \includegraphics[width=0.65\linewidth]{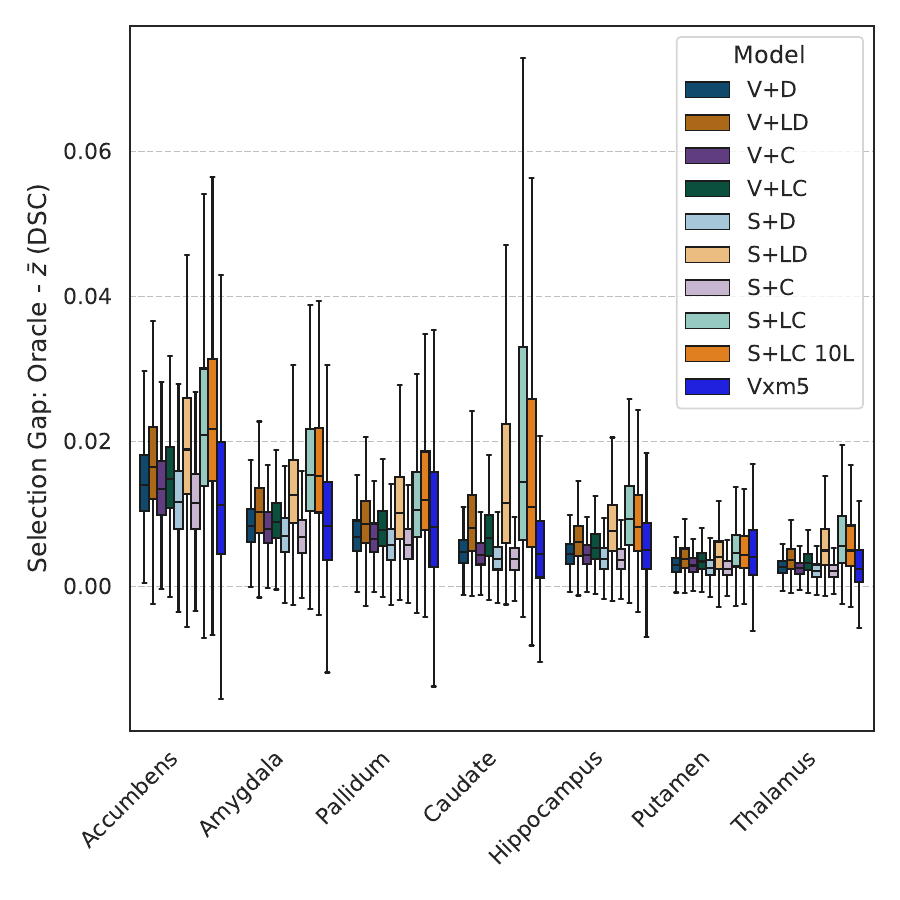}
    \vspace{-0.5cm}
    \caption{Boxplot highlighting the quality of the Oracle sample drawn from the approximate posterior distribution for the full set of segmentation regions. Larger Dice Similarity Coefficient (DSC) values indicated greater segmentation accuracy.}
\end{figure}

\FloatBarrier\newpage

\section{Calibration}
\label{sec:calibration_app}
\begin{figure}[ht!]
    \centering
    \includegraphics[width=0.9\linewidth]{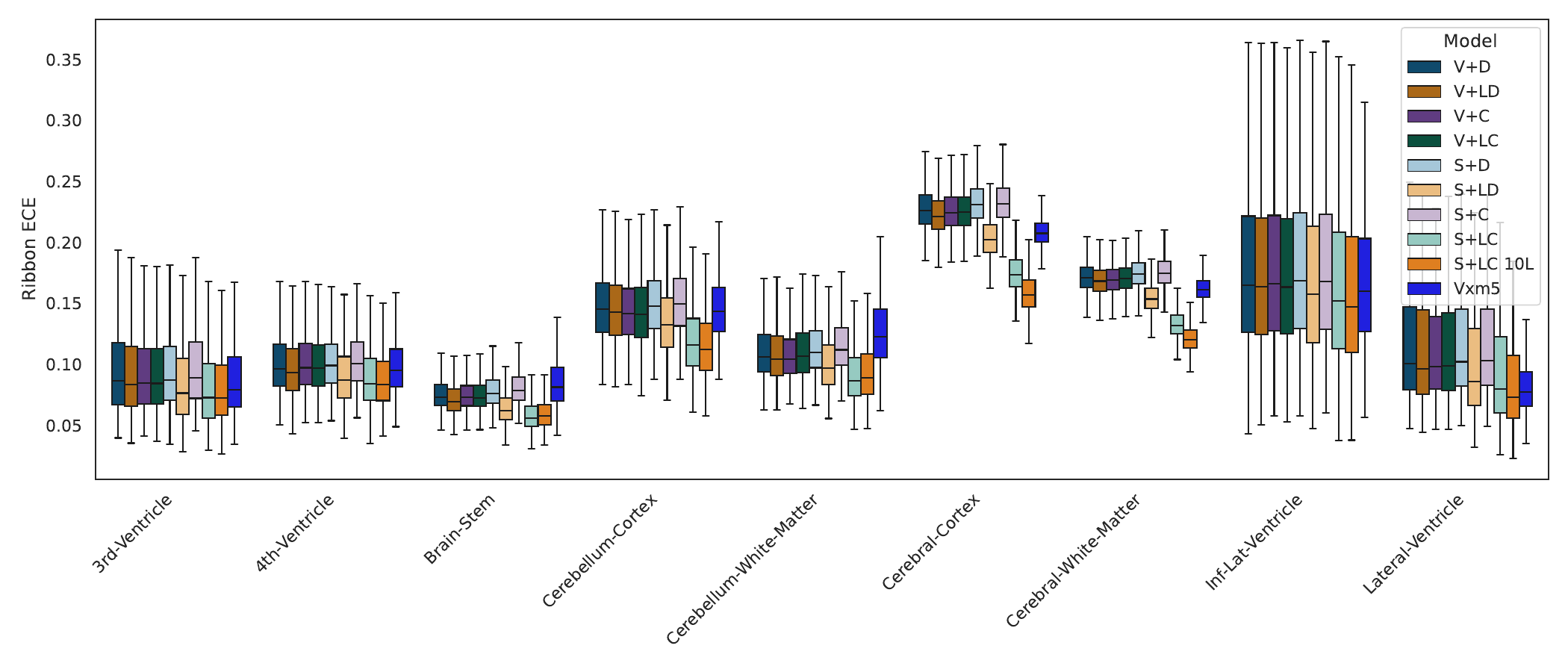}
    \includegraphics[width=0.9\linewidth]{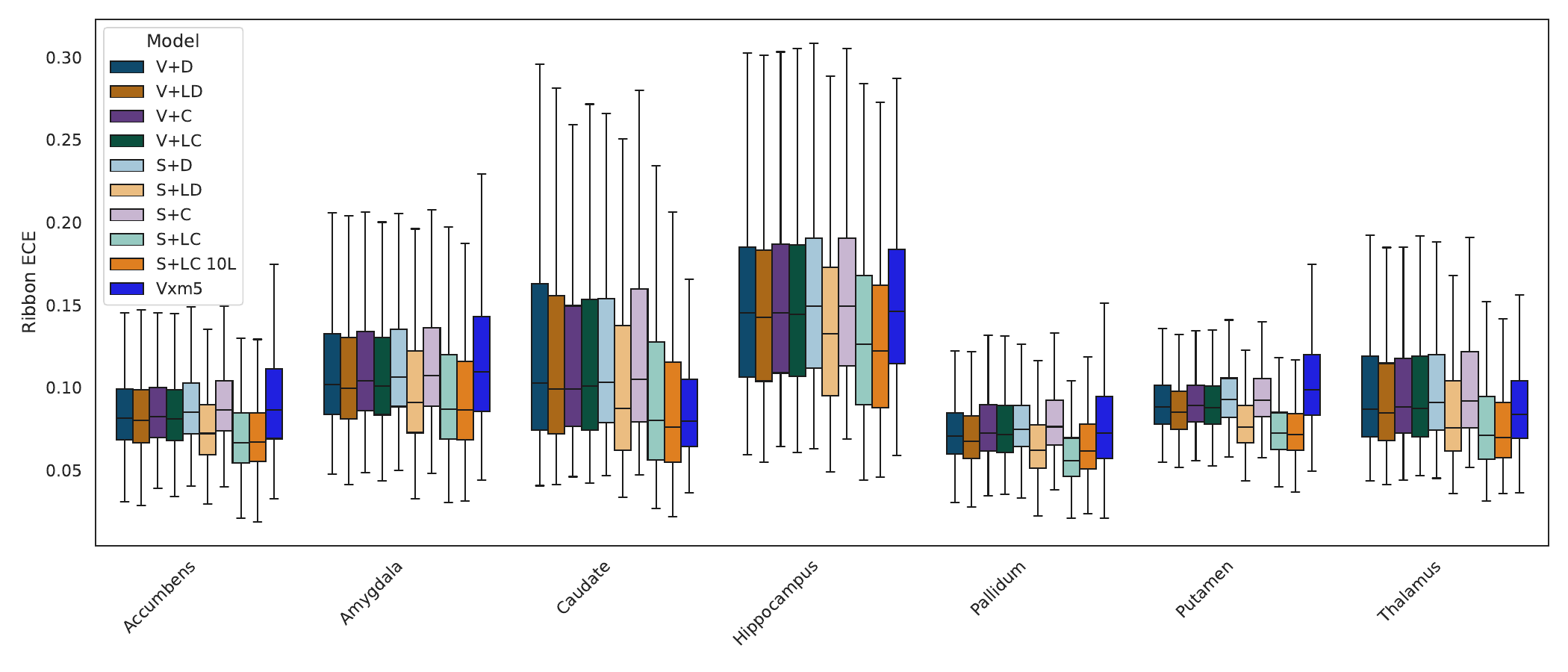}
    \caption{Expected calibration error (ECE) boxplots across all model variants and structures. ECE was calculated in a region defined by a 3 voxels dilated segmentation mask.}
    \label{fig:ece_all}
\end{figure}
\begin{figure}[ht!]
    \centering
    \includegraphics[width=0.9\linewidth]{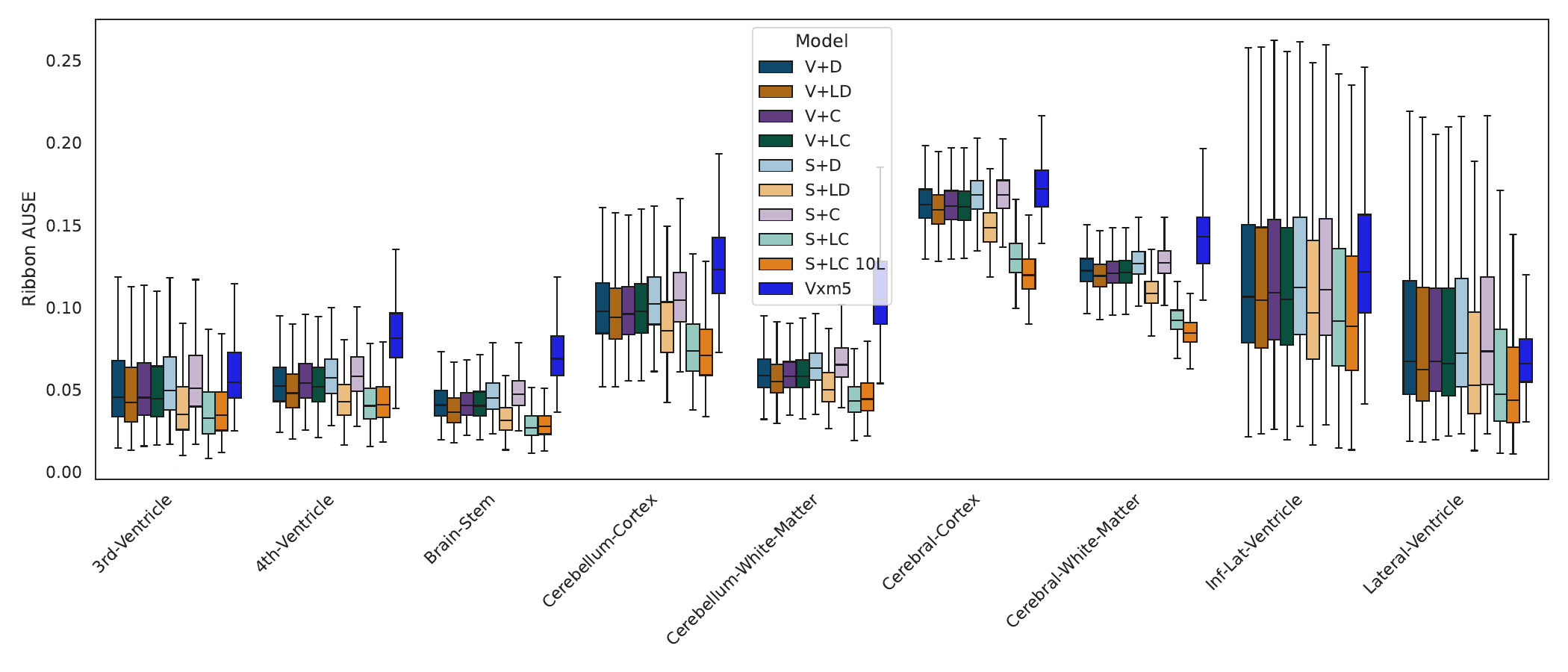}
    \includegraphics[width=0.9\linewidth]{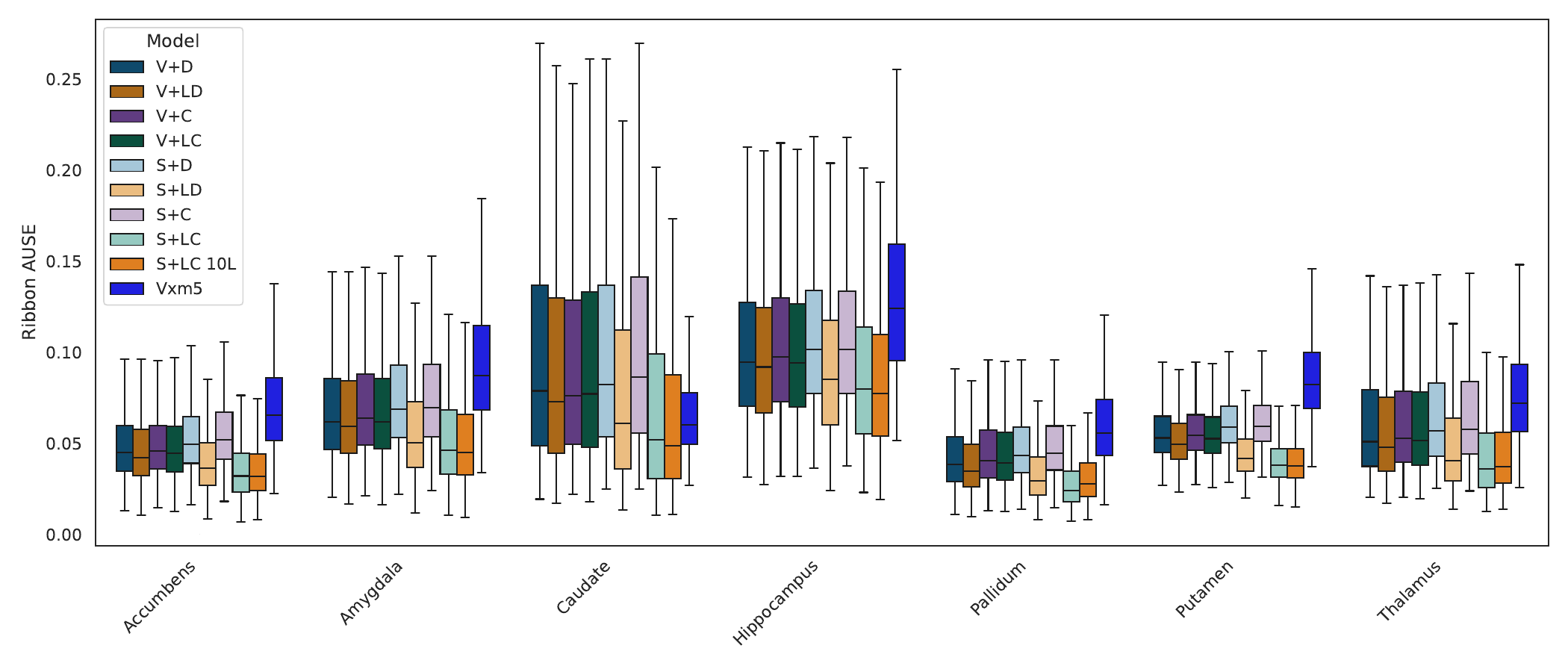}
    \caption{Area Under Sparse Error (AUSE) boxplots across all model variants and structures. AUSE was calculated in a region defined by a 3 voxels dilated segmentation mask using the label entropy.}
    \label{fig:ause_all}
\end{figure}

We provide boxplots showing the calibration metrics across all model variance and structures in \cref{fig:ece_all} and \cref{fig:ause_all}

In \cref{tab:ause_disp} shows a companion representation to \cref{tab:calibration_metrics} where we instead use displacement entropy for calculating AUSE, and correlating with the Dice score. 
Here we find that displacement entropy is negatively correlated with Dice for all structures, although the level of entropy is much lower for all except our best calibrated model variant. 
While displacement entropy quantifies the geometric variance of the deformation field, it captures invisible jitter in homogeneous regions that does not impact anatomical overlap; in contrast, label entropy directly measures the probabilistic risk of misclassification at structural boundaries, providing a more anatomically relevant measure of uncertainty.

\begin{table}[h!]
\caption{Uncertainty Calibration Metrics: Area Under Sparse Error (AUSE) based on displacement entropy  calculated using a dilated (by 3 voxels) segmentation mask. The last column is the Spearman correlation of label entropy and Dice. All fields show the average and std-dev over structures.}
\label{tab:ause_disp}
\begin{tabular}{l|cc c }
\toprule
 & $\mathrm{AUSE}_{\mathrm{disp}} \downarrow$ & Displacement Entropy $(\mathrm{mm}^2)$ & Spearman r: Entropy \& DSC \\
Model &  &  &  \\
\midrule
V+D & 0.1351 (0.017) & 0.0494 (0.001) & -0.648 (0.121) \\
V+LD & 0.1343 (0.017) & 0.0635 (0.002) & -0.430 (0.168) \\
V+C & 0.1312 (0.015) & 0.0449 (0.001) & -0.628 (0.138) \\
V+LC & 0.1323 (0.015) & 0.0571 (0.005) & -0.530 (0.221) \\
S+D & 0.1366 (0.018) & 0.0432 (0.006) & -0.199 (0.059) \\
S+LD & 0.1272 (0.017) & 0.1128 (0.014) & -0.549 (0.166) \\
S+C & 0.1339 (0.017) & 0.0403 (0.005) & -0.311 (0.099) \\
S+LC & 0.1205 (0.016) & 0.1808 (0.037) & -0.595 (0.156) \\
S+LC 10L & 0.1186 (0.016) & 0.2270 (0.074) & -0.624 (0.165) \\
Vxm & 0.1223 (0.014) & 0.1428 (0.0909) & -0.571 (0.199) \\
\bottomrule
\end{tabular}
\end{table}

\newpage
\section{Training Stability}\label{sec:stability}
\begin{figure}[h]
    \includegraphics[width=.32\linewidth]{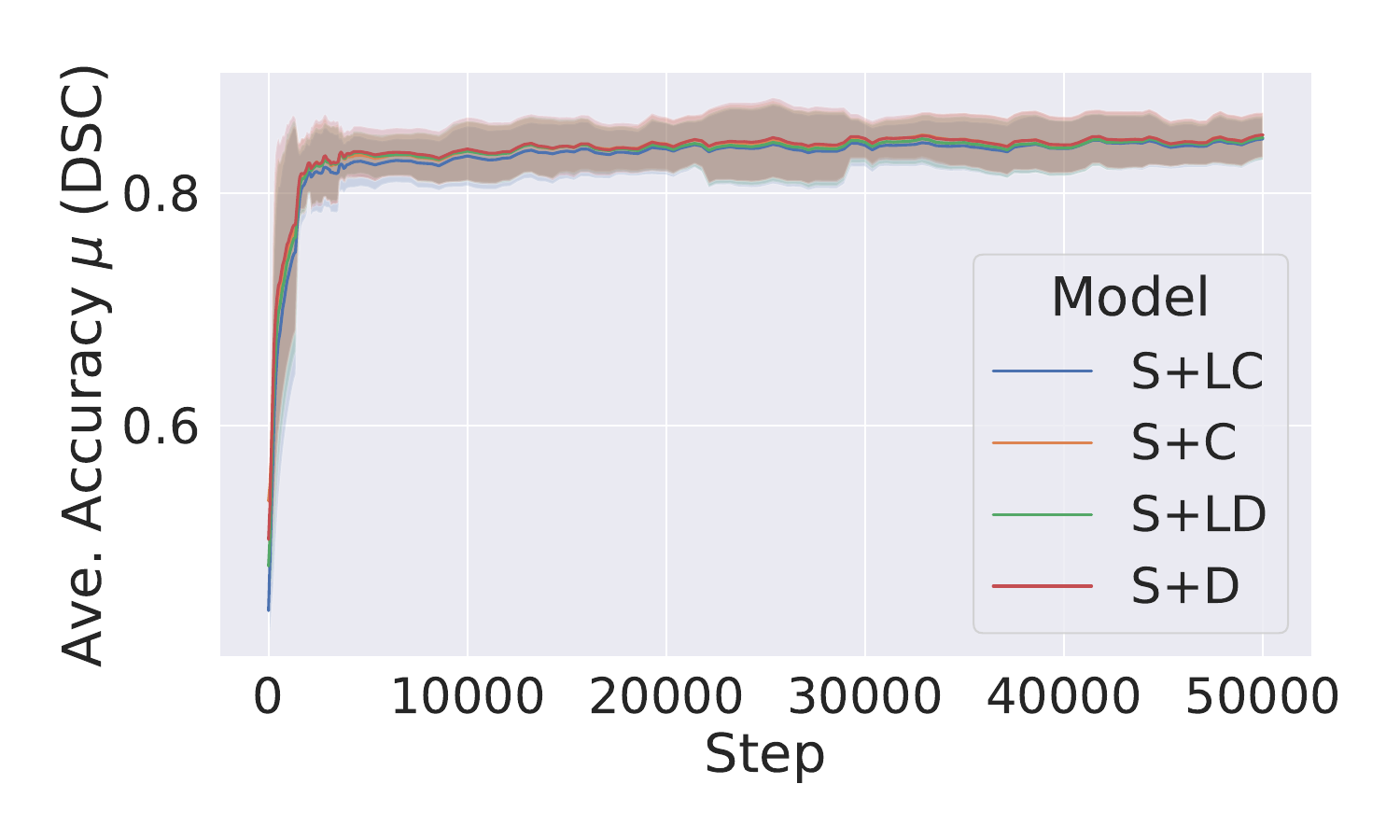}
    \includegraphics[width=.32\linewidth]{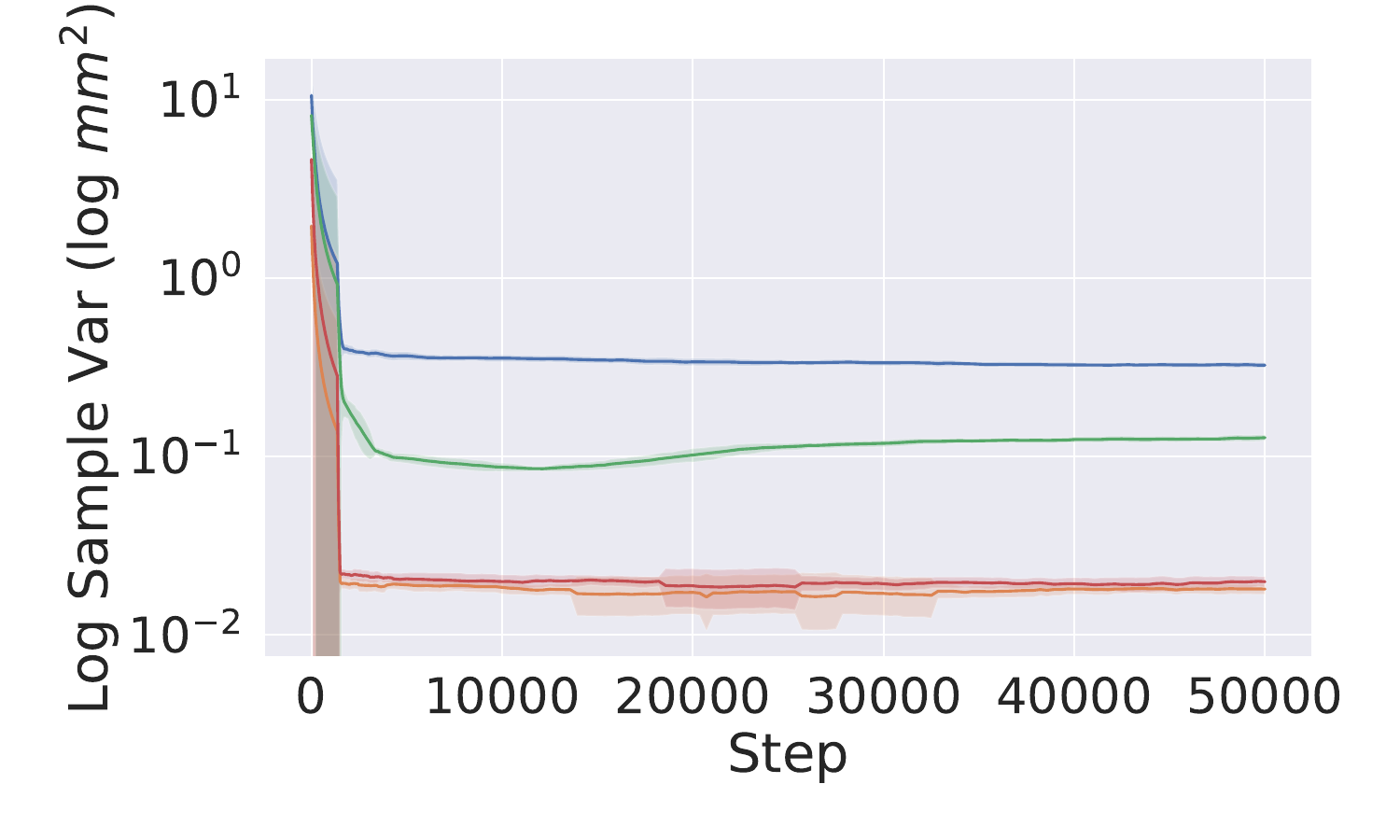}
    \includegraphics[width=.32\linewidth]{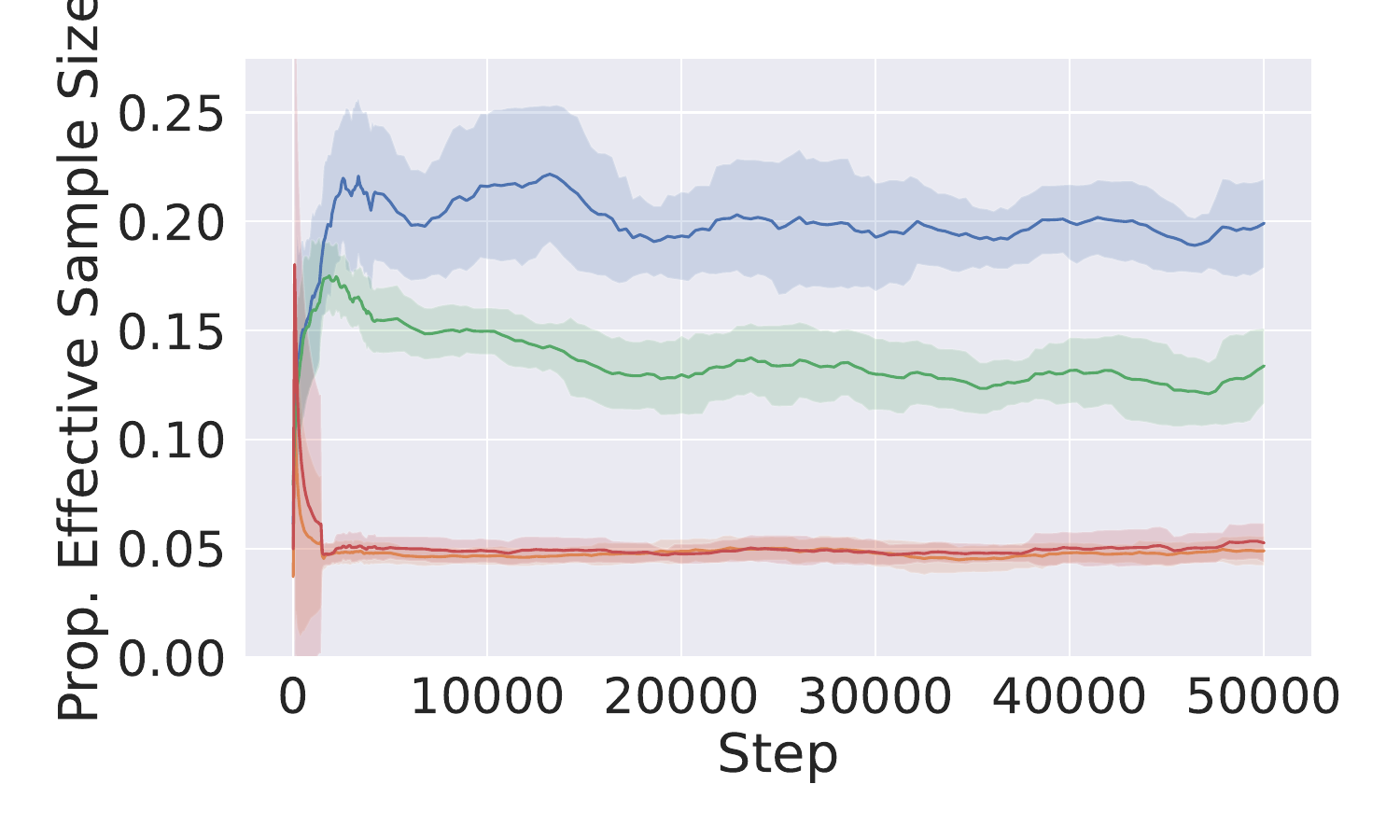}
    \caption{Average accuracy (left), displacement variance (mid) and proportional effective sample size ($1.0 / \sum_i w_i ^ 2$, right) on the val set.
    }
    \label{fig:training_stability}
\end{figure}

\cref{fig:training_stability} demonstrates that accuracy, displacement variance and proportional effective sample size (ESS) stabilize smoothly from the cold start, refuting weight degeneracy and confirming that loss signals reliably steer poor early samples.

\newpage
\section{Sensitivity Analysis}

\subsection{Number of low-rank components}\label{sec:app_num_low_rank}
\begin{table}
\caption{Registration performance and folding rates across OASIS test pairs modifying the number of low-rank components, with all other hyperparameters held constant. We report the Dice Similarity Coefficient (DSC) for the mean prediction ($\mu_{\warpp}$), the mean of resampled fields ($\bar{\warpp}$), and the Oracle (the single sample from the posterior with the highest DSC). Fold \% represents the percentage of voxels with a non-positive Jacobian determinant. Standard deviations are shown in parentheses. High DSC values and lower fold \% are preferable.}
\label{tab:model_comparison_lr}
\begin{tabular}{l|ccc|ccc}
\toprule
 & $\qmean$ (DSC) & $\bar{\warpp}$ (DSC) & Oracle (DSC) & $\qmean$ fold \% & $\bar{\warpp}$ fold \% & Oracle fold \% \\
Model &  &  &  &  &  &  \\
\midrule
S+LC & 0.779 (0.05) & 0.804 (0.04) & 0.815 (0.03) & 0.001 (0.00) & 0.453 (0.18) & 0.458 (0.36) \\
5L &  &  &  &  &  &   \\
S+LC & 0.788 (0.04) & 0.802 (0.04) & 0.814 (0.03) & 0.004 (0.01) & 0.240 (0.14) & 0.292 (0.25) \\
10L &  &  &  &  &  &   \\
S+LC & 0.794 (0.04) & 0.801 (0.04) & 0.812 (0.03) & 0.017 (0.01) & 0.143 (0.11) & 0.201 (0.18) \\
20L &  &  &  &  &  &   \\
S+LC & 0.794 (0.04) & 0.801 (0.04) & 0.812 (0.03) & 0.011 (0.01) & 0.115 (0.07) & 0.163 (0.13) \\
25L &  &  &  &  &  &   \\
S+LC & 0.796 (0.04) & 0.800 (0.04) & 0.810 (0.04) & 0.024 (0.02) & 0.092 (0.07) & 0.116 (0.10) \\
30L &  &  &  &  &  &   \\
\bottomrule
\end{tabular}
\end{table}

\begin{table}
\caption{Uncertainty Calibration Metrics comparing the number of low-rank components, with all other hyperparameters held constant. Area Under Sparse Error (AUSE) based on label entropy and Expected Calibration Error (ECE) calculated using a dilated (by 3 voxels) segmentation mask. The last column is the Spearman correlation of label entropy and Dice. All fields show the average and std-dev over structures.}
\label{tab:calibration_metrics_lr}
\begin{tabular}{l|p{22mm} p{22mm} p{22mm} p{27mm}}
\toprule
 & $AUSE_{label} \downarrow$ & $ECE \downarrow$ & Label Entropy & Mean Spearman r: Entropy \& DSC \\
Model &  &  &  &  \\
\midrule
S+LC 5L& 0.0646 (0.018) & 0.1030 (0.022) & 0.1128 (0.011) & -0.400 (0.224) \\
S+LC 10L& 0.0636 (0.018) & 0.1028 (0.023) & 0.1156 (0.010) & -0.419 (0.217) \\
S+LC 20L& 0.0681 (0.019) & 0.1087 (0.024) & 0.1046 (0.007) & -0.376 (0.221) \\
S+LC 25L & 0.0659 (0.019) & 0.1064 (0.025) & 0.1079 (0.006) & -0.325 (0.218) \\
S+LC 30L& 0.0725 (0.020) & 0.1143 (0.026) & 0.0940 (0.005) & -0.241 (0.194) \\
\bottomrule
\end{tabular}
\end{table}

We evaluate the impact of the number of low-rank components in our S+LC approach in \cref{tab:model_comparison_lr} and \cref{tab:calibration_metrics_lr}. We find that 10 components provides an optimal number for uncertainty calibration. However, as all other experiments are conducted with 25 components we report this as our main result.

\subsection{Temperature scaling}\label{sec:app_temperature}
We provide results for adjusting the Temperature value $T$ in \cref{tab:model_comparison_temp} and \cref{tab:calibration_metrics_temp}. We note that there may be small gains in further optimise the temperature choice.

\begin{table}
\caption{Registration performance and folding rates across OASIS test pairs modifying temperature values $T$, with all other hyperparameters held constant. We report the Dice Similarity Coefficient (DSC) for the mean prediction ($\mu_{\warpp}$), the mean of resampled fields ($\bar{\warpp}$), and the Oracle (the single sample from the posterior with the highest DSC). Fold \% represents the percentage of voxels with a non-positive Jacobian determinant. Standard deviations are shown in parentheses. High DSC values and lower fold \% are preferable.}
\label{tab:model_comparison_temp}
\begin{tabular}{l|ccc|ccc}
\toprule
 & $\qmean$ (DSC) & $\bar{\warpp}$ (DSC) & Oracle (DSC) & $\qmean$ fold \% & $\bar{\warpp}$ fold \% & Oracle fold \% \\
Model &  &  &  &  &  &  \\
\midrule
S+LC & 0.800 (0.04) & 0.801 (0.04) & 0.809 (0.04) & 0.043 (0.04) & 0.073 (0.07) & 0.079 (0.08) \\ 
$T=2.0$  &  &  &  &  &  & \\
S+LC & 0.794 (0.04) & 0.801 (0.04) & 0.812 (0.03) & 0.011 (0.01) & 0.115 (0.07) & 0.163 (0.13) \\
$T=3.0$  &  &  &  &  &  & \\
S+LC & 0.792 (0.04) & 0.802 (0.04) & 0.813 (0.03) & 0.009 (0.01) & 0.185 (0.12) & 0.224 (0.19) \\
$T=4.0$  &  &  &  &  &  & \\
\bottomrule
\end{tabular}
\end{table}

\begin{table}
\caption{Uncertainty Calibration Metrics across temperature values $T$, with all under parameters held constant: Area Under Sparse Error (AUSE) based on label entropy and Expected Calibration Error (ECE) calculated using a dilated (by 3 voxels) segmentation mask. The last column is the Spearman correlation of label entropy and Dice. All fields show the average and std-dev over structures.}
\label{tab:calibration_metrics_temp}
\begin{tabular}{l|p{22mm} p{22mm} p{22mm} p{27mm}}
\toprule
 & $AUSE_{label} \downarrow$ & $ECE \downarrow$ & Label Entropy & Mean Spearman r: Entropy \& DSC \\
Model &  &  &  &  \\
\midrule
S+LC $T=2.0$ & 0.0775 (0.020) & 0.1207 (0.026) & 0.0780 (0.003) & -0.119 (0.146) \\
S+LC $T=3.0$ & 0.0659 (0.019) & 0.1064 (0.025) & 0.1079 (0.006) & -0.325 (0.218) \\
S+LC $T=4.0$ & 0.0650 (0.018) & 0.1044 (0.023) & 0.1114 (0.009) & -0.408 (0.219) \\
\bottomrule
\end{tabular}
\end{table}

\newpage
\section{Baseline Model Details}
\label{sec:app_baselines}
We compare our approach against a deep ensemble of 5 Voxelmorph \citep{dalca2019unsupervised} models. These models were trained with mean-square error, diffusion regularisation, batch size 4 and $\lambda=0.01$ with equivalent model complexity to ours. Fields were produced at half the image size and upsampled with cubic interpolation, which matches our approach. We also compare against a resolution matched deterministic method, Convex Adam \citep{heinrich2014non} with MIND features \citep{heinrich2012mind}, and also a full-resolution Voxelmorph ensemble. We present a brief summary of the results in \cref{tab:baseline_comparison} and note that the full resolution Voxelmorph has unsurprisingly higher accuracy, but still suffers from worse label AUSE (with similar ECE). SIR based models could also be tractably trained and run at full resolution given sufficient computational resource.

\begin{table}
\caption{Comparison of model variants with Convex Adam (CA) with MIND features, a Voxelmorph 5-model ensemble at matched resolution (Vxm5) and a Voxelmorph 5-model ensemble at maximum resolution} 
\label{tab:baseline_comparison}
\begin{tabular}{|l|ccc| c c c |}
\toprule
 & $\qmean$ (DSC) & $\bar{\warpp}$ (DSC) & Oracle (DSC) & $AUSE_{label} \downarrow$ & $ECE \downarrow$ & Label Entropy   \\
\midrule
V+D & 0.800 (0.04) & 0.800 (0.04) & 0.805 (0.04) & 0.085 (0.02) & 0.129 (0.03) & 0.063 (0.00) \\
S+LD & 0.798 (0.04) & 0.800 (0.04) & 0.809 (0.04) & 0.074 (0.02) & 0.116 (0.03) & 0.090 (0.00) \\
S+LC & 0.794 (0.04) & 0.801 (0.04) & 0.812 (0.03)) & 0.066 (0.02) & 0.107 (0.03) & 0.108 (0.01)\\
S+LC & 0.788 (0.04) & 0.802 (0.04) & 0.814 (0.03) &  0.064 (0.02) & 0.103 (0.02) & 0.115 (0.01) \\
10L & & & & & & \\
\midrule
CA & 0.788 (0.04) & 0.788 (0.04) & 0.788 (0.04) & - & -& -\\
\midrule
Vxm5 & 0.798 (0.03) & 0.798 (0.03) & 0.804 (0.03) & 0.101 (0.02) & 0.122 (0.02) & 0.070 (0.01)  \\
\midrule
\midrule
Vxm5 & 0.818 (0.03) & 0.818 (0.03) & 0.823 (0.03) & 0.081 (0.02) & 0.099 (0.02) & 0.079 (0.01) \\
max & & & & & &\\
\bottomrule
\end{tabular}
\end{table}

\newpage
\section{Uncertainty Metric Details: ECE and AUSE}
\label{app:uncertainty_metrics}

To evaluate the quality of the predicted spatial uncertainty, we compute the Expected Calibration Error (ECE) and Area Under the Sparsification Error (AUSE) over a localised spatial domain where registration ambiguity is likely to be most prevalent. 

\subsection{Domain Localization and Probability Mapping}
Let $\Omega$ denote the volumetric image domain. We define the anatomical ribbon $\Omega_R \subset \Omega$ via a 3-voxel morphological dilation of the ground-truth anatomical label boundaries, containing $N = |\Omega_R|$ evaluation voxels. For each voxel $x \in \Omega_R$, the empirical posterior probability $\hat{p}(x)$ of a given anatomical structure is computed as the average across $K$ warped label samples:
\begin{equation}
\hat{p}(x) = \frac{1}{K} \sum_{k=1}^K Y_k(x)
\end{equation}
where $Y_k(x) \in \{0, 1\}$ is the propagated binary label for sample $k$. The binary voxel error map is given by $E(x) = |\mathbb{I}(\hat{p}(x) > 0.5) - Y_{GT}(x)| \in \{0, 1\}$, where $Y_{GT}(x)$ is the ground-truth target label and $\mathbb{I}$ is the indicator function.

\subsection{Expected Calibration Error (ECE)}
ECE measures the absolute discrepancy between the model's confidence and its actual empirical accuracy. The voxels within $\Omega_R$ are partitioned into $M$ equally spaced confidence bins $B_m = (\frac{m-1}{M}, \frac{m}{M}]$ based on $\hat{p}(x)$. The ECE is computed as a weighted average across all bins:
\begin{equation}
\text{ECE} = \sum_{m=1}^M \frac{|B_m|}{N} \Big| \text{conf}(B_m) - \text{acc}(B_m) \Big|
\end{equation}
where the mean confidence $\text{conf}(B_m)$ and empirical accuracy $\text{acc}(B_m)$ of bin $B_m$ are defined as:
\begin{equation}
\text{conf}(B_m) = \frac{1}{|B_m|} \sum_{x \in B_m} \hat{p}(x), \quad \text{acc}(B_m) = \frac{1}{|B_m|} \sum_{x \in B_m} Y_{GT}(x)
\end{equation}

\subsection{Area Under the Sparsification Error (AUSE)}
AUSE assesses the error-ranking capability of an uncertainty metric $u(x)$ (evaluated independently for displacement variance and label entropy). Voxels in $\Omega_R$ are permuted via two separate ordering functions:
\begin{itemize}
    \item \textbf{Observed Sorting ($\pi_{unc}$):} Ordered by descending uncertainty: $u(\pi_{unc}(1)) \ge u(\pi_{unc}(2)) \ge \dots \ge u(\pi_{unc}(N))$.
    \item \textbf{Oracle Sorting ($\pi_{err}$):} Ordered by descending true error: $E(\pi_{err}(1)) \ge E(\pi_{err}(2)) \ge \dots \ge E(\pi_{err}(N))$.
\end{itemize}
Sparsification curves $c_{obs}(i)$ and $c_{ora}(i)$ track the remaining mean error after sequentially removing the highest $N-i$ entries according to their respective permutations:
\begin{equation}
c_{obs}(i) = \frac{1}{i} \sum_{j=1}^i E(\pi_{unc}(j)), \quad c_{ora}(i) = \frac{1}{i} \sum_{j=1}^i E(\pi_{err}(j))
\end{equation}
The AUSE integrates the gap between the two curves across all remaining voxel fractions $i \in \{1, \dots, N\}$:
\begin{equation}
\text{AUSE} = \frac{1}{N} \sum_{i=1}^N \left( c_{obs}(i) - c_{ora}(i) \right)
\end{equation}
An AUSE of 0 denotes perfect error-ranking capability, demonstrating that the model's localised variance directly identifies registration failures.

\end{document}